\documentclass[a4paper,11pt]{article}
\pdfoutput=1
\usepackage{jheppub}
\usepackage[T1]{fontenc}
\usepackage{amsthm}
\usepackage{tikz}

\theoremstyle{plain}
\newtheorem{thm}{Theorem}[section]
\newtheorem{lem}[thm]{Lemma}
\newtheorem{prop}[thm]{Proposition}

\theoremstyle{definition}
\newtheorem{defn}[thm]{Definition}
\newtheorem{exmp}[thm]{Example}
\newtheorem{rem}[thm]{Remark}

\DeclareMathOperator{\conf}{conf}
\DeclareMathOperator{\diff}{Diff}

\title{Dynamics for holographic codes}

\author[a]{Tobias J. Osborne,}
\author[a,b]{Deniz E. Stiegemann}

\affiliation[a]{Institut f\"ur Theoretische Physik, Leibniz Universit\"at Hannover, Appelstra{\ss}e 2, 30167 Hannover, Germany}
\affiliation[b]{ARC Centre for Engineered Quantum Systems, School of Mathematics and Physics,
The University of Queensland, Brisbane, QLD 4072, Australia}

\emailAdd{tobias.osborne@itp.uni-hannover.de}
\emailAdd{deniz@stiegemann.com}

\abstract{We describe how to introduce dynamics for the holographic states and codes introduced by Pastawski, Yoshida, Harlow and Preskill. This task requires the definition of a continuous limit of the kinematical Hilbert space which we argue may be achieved via the \emph{semicontinuous limit} of Jones. Dynamics is then introduced by building a unitary representation of a group known as Thompson's group $T$, which is closely related to the conformal group $\conf(\mathbb{R}^{1,1})$. The bulk Hilbert space is realised as a special subspace of the semicontinuous limit Hilbert space spanned by a class of distinguished states which can be assigned a discrete bulk geometry. The analogue of the group of large bulk diffeomorphisms is given by a unitary representation of the \emph{Ptolemy group} $\mathit{Pt}$, on the bulk Hilbert space thus realising a toy model of the AdS/CFT correspondence which we call the $\mathit{Pt}/T$ correspondence.}

\begin{document} 
\maketitle
\flushbottom

\section{Intoduction}

The phenomenon of \emph{holographic duality}, whereby a strongly interacting boundary quantum field theory is dual to a bulk quantum gravity theory in the semiclassical limit (and vice versa) has lead to an extraordinarily powerful way to reason about quantum gravity theories. This is due in no small part to the work of Maldacena \cite{maldacena_large-n_1999,maldacena_large_1998} who first found a quantitative argument that there is an equivalence between string theory on $\text{AdS}_5\times S^5$ and $\mathcal{N}=4$ supersymmetric Yang--Mills theory on the four-dimensional boundary. Maldacena's papers generated an explosion of work consolidating and exploring holographic dualities, starting with the first works of Gubser, Klebanov, and Polyakov \cite{gubser_gauge_1998}, and Witten \cite{witten_anti_1998} and continuing unabated to the current day.

One striking result of the AdS/CFT correspondence is the duality between bulk geometry and the pattern of quantum entanglement in the boundary theory. This connection manifests itself in the formula of Ryu and Takayanagi \cite{ryu_holographic_2006,rangamani_holographic_2016} expressing the entropy of a region in the boundary CFT with the area of a specific bulk minimal surface. This observation has been considerably expanded and developed in recent years commencing with the proposals of Van Raamsdonk \cite{van_raamsdonk_building_2010} and Swingle \cite{swingle_entanglement_2012} and strengthened by the $\text{ER}=\text{EPR}$ proposal of Susskind and Maldacena \cite{maldacena_cool_2013}.  These ideas have recently sparked a very fertile line of enquiry in which results from quantum information theory developed to quantify quantum entanglement and Hamiltonian complexity theory are used to understand the quantum dynamics of black holes \cite{harlow_jerusalem_2016,brown_complexity_2016}.

Quantum information theory is also playing an increasingly important role in the study of holographic duality. For example, it was realised that new techniques would be required to address some apparantly paradoxical features of the bulk/boundary correspondence, in particular, in understanding the boundary dual of local bulk operators. In a prescient and influential work, Almheiri, Harlow, and Dong \cite{almheiri_bulk_2015} exploited the theory of quantum error correction to resolve these ambiguities. They argue that bulk local operators should manifest themselves as logical operators on subspaces of the boundary CFT's $\mathrm{AdS}_3$ space.

The connection between quantum error correction and the AdS/CFT correspondence, as highlighted by Almheiri, Harlow, and Dong, was dramatically illustrated by the construction of a beautiful toy model of holographic duality known as the \emph{holographic code} \cite{pastawski_holographic_2015}. This is a discrete model of the kinematical content of the AdS/CFT correspondence built on special tensors known as \emph{perfect tensors} arising from certain quantum error correcting codes. The holographic code has proved extremely helpful as a testbed for conjectures and also as a sandbox for refinements of the holographic dictionary. In the two years since its inception there have been numerous papers investigating and generalising both holographic codes \cite{pastawski_code_2016,hayden_holographic_2016,bao_consistency_2015,yang_bidirectional_2016,bhattacharyya_exploring_2016,may_tensor_2016}
 and perfect tensors \cite{goyeneche_absolutely_2015,enriquez_maximally_2016,raissi_constructing_2017,li_invariant_2016,peach_tensor_2017,donnelly_living_2016}. Noticing that holographic codes are tensor networks with a particular causal structure, it is tempting to hope that there is a more general manifestation of the AdS/CFT correspondence arising from the network structure alone \cite{swingle_entanglement_2012,czech_integral_2015,beny_causal_2013,czech_tensor_2016,qi_exact_2013}.

One key question is largely open in the context of realising a full \emph{dynamical} toy model of the AdS/CFT correspondence via quantum codes and perfect tensors: How can one realise \emph{dynamics} for holographic codes? It is this question that we address in this paper in the context of the simplest possible holographic code, namely the holographic state coming from a tree tensor network built with a 3-leg perfect tensor. To do this we leverage powerful new results \cite{jones_unitary_2014,jones_no-go_2016} of Jones on unitary representations of a discrete analogue of the conformal group known as Thompson's group $T$ \cite{cannon_introductory_1996,belk_thompsons_2007}. We argue that dynamics for the holographic state should be given precisely as a unitary representation of $T$. It is worth emphasising that this approach is distinct from the $p$-adic AdS/CFT correspondence \cite{gubser_p-adic_2017,heydeman_tensor_2016} (see also \cite{harlow_tree-like_2012}) as the group of symmetries in this case is not isomorphic to $T$.

This paper is largely synthesised from a great deal of hard work done by Penner, Funar, Sergiescu, and Jones \cite{schneps_geometric_1997,funar_central_2010,jones_unitary_2014,jones_no-go_2016,jones_scale_2017}. We make no claim on the originality of the unitary representations described herein, nor on the connection between Thompson's group $T$ and the Ptolemy group(oid). The main contribution of this work is to notice that when Penner's work on the Ptolemy group(oid) is combined with Jones's unitary representations via a holographic state, we get a dynamical toy model of the AdS/CFT correspondence. We also discuss how one could generalise these results to holographic states built on more general tessellations of the hyperbolic plane.

\section{The building blocks}

In this section we explain how to form a Hilbert space for states that are, in a suitable sense, rotation-invariant on a discrete approximation of $\mathrm{AdS}_3$.

\subsection{Perfect tensors}

The notion of a perfect tensor was introduced in \cite{pastawski_holographic_2015}. These highly nongeneric objects capture a discrete version of rotation invariance which is extremely useful in building network approximations to continuous manifolds.

We employ a tensor-network representation for $n$-index tensors: Suppose that $T_{j_1j_2\cdots j_n}$ is a tensor with $n$ indices, each ranging from $0$ to $d-1$. Then $T$ is depicted as a vertex with $d$ legs, where each leg represents one of the indices in counter-clockwise order (figure~\ref{fig:tensornotation}). By convention, the first leg, here indexed by $j_1$, is the one directly following the label ``$T\,$'' in counter-clockwise order.
\begin{figure}
  \centering
  \begin{tikzpicture}
    \foreach \i in {1, ..., 8}{
      \draw (0, 0) -- ({(\i-1)*360/8}:1.6) node[pos=1.2] {$j_\i$};
    }
    \draw[fill=white] (0, 0) circle[radius=0.4] node {$T$};

    \begin{scope}[xshift=6cm]
      \foreach \i in {1, ..., 8}{
        \draw (0, 0) -- ++({(\i-1)*360/8}:1.6);
      }
      \draw[fill] (0, 0) circle[radius=0.1];
      \node at (-15:1) {$T$};
    \end{scope}
  \end{tikzpicture}
  \caption{In the usual tensor network notation, a tensor $T$ with, say, seven indices is depicted as a node with seven legs (left diagram). We will follow the convention that the first leg is the one following the name of the tensor (here $T$) in counter-clockwise order.}
  \label{fig:tensornotation}
\end{figure}

We now come to perfect tensors.
\begin{defn}
An $n$-index tensor $T_{j_1j_2\ldots j_n}$ is a \emph{perfect tensor} if, for any bipartition of its indices into a pair of complementary sets $\{j_1, j_2, \ldots, j_n\}=A\cup A^c$ such that, without loss of generality, $|A|\le |A^c|$, $T$ is proportional to an isometry from the Hilbert space associated with $A$ to the Hilbert space associated with $A^c$.
\end{defn}
If the legs have dimension $d$ then, for a given partition $\{j_1, j_2, \ldots, j_n\}= A\cup A^c$, $T$ is a linear map
\begin{equation}
	T\colon\bigotimes_{j\in A} \mathbb{C}^d \to \bigotimes_{j\in A^c} \mathbb{C}^d.
\end{equation}
In particular, if $A = \emptyset$ and $A^c = \{j_1, j_2, \ldots, j_n\}$, then $T$ is a vector in the Hilbert space $({\mathbb{C}^d})^{\otimes n}$. From now on we write $d$ for the (constant) dimension of the legs of a perfect tensor.

The perfect tensor condition is nontrivial and it is far from obvious whether perfect tensors exist at all (they do). Throughout this paper we  focus on the simplest case of $n=3$. In this particularly simple setting the definition of a perfect tensor reduces to that of a $3$-leg tensor $V$ which is an isometry in all three possible directions, as depicted in figure~\ref{fig:perfectcondition}.
\begin{figure}
  \centering
  \begin{tikzpicture}
    \foreach \i in {0.75, 1.75, 2.75}{
      \draw (0, 0) -- ++({\i*360/3}:1.3);
    }
    \draw[fill] (0, 0) circle[radius=0.1];
    \node at (65:0.7) {$V$};
    \begin{scope}[yshift=-1.3cm]
      \foreach \i in {0.25, 1.25, 2.25}{
        \draw (0, 0) -- ++({\i*360/3}:1.3);
      }
      \draw[fill] (0, 0) circle[radius=0.1];
      \node at (245:0.7) {$V^\dagger$};
    \end{scope}
    \node at (1.5, -0.65) {$=$};
    \draw (2, 1.3) -- (2, -2.6);

    \begin{scope}[xshift=6cm]
      \foreach \i in {0.75, 1.75, 2.75}{
        \draw (0, 0) -- ++({\i*360/3}:1.3);
      }
      \draw[fill] (0, 0) circle[radius=0.1];
      \node at (-55:0.7) {$V$};
      \begin{scope}[yshift=-1.3cm]
        \foreach \i in {0.25, 1.25, 2.25}{
          \draw (0, 0) -- ++({\i*360/3}:1.3);
        }
        \draw[fill] (0, 0) circle[radius=0.1];
        \node at (5:0.7) {$V^\dagger$};
      \end{scope}
      \node at (1.5, -0.65) {$=$};
      \draw (2, 1.3) -- (2, -2.6);
    \end{scope}

    \begin{scope}[xshift=3cm, yshift=-4.5cm]
      \foreach \i in {0.75, 1.75, 2.75}{
        \draw (0, 0) -- ++({\i*360/3}:1.3);
      }
      \draw[fill] (0, 0) circle[radius=0.1];
      \node at (185:0.7) {$V$};
      \begin{scope}[yshift=-1.3cm]
        \foreach \i in {0.25, 1.25, 2.25}{
          \draw (0, 0) -- ++({\i*360/3}:1.3);
        }
        \draw[fill] (0, 0) circle[radius=0.1];
        \node at (125:0.7) {$V^\dagger$};
      \end{scope}
      \node at (1.5, -0.65) {$=$};
      \draw (2, 1.3) -- (2, -2.6);
    \end{scope}
  \end{tikzpicture}
  \caption{The conditions for a $3$-leg tensor $V$ to be perfect. The straight line represents the identity.}
  \label{fig:perfectcondition}
\end{figure}
 
Here are some interesting examples of perfect tensors to keep in mind in the sequel.
\begin{exmp}
  For the case $n=2$ any unitary $U$ is perfect.
\end{exmp}
\begin{exmp}
  Let $n=3$ and $d =4$ and define the map $V\colon\mathbb{C}^2\otimes \mathbb{C}^2 \to \mathbb{C}^2\otimes\mathbb{C}^2\otimes\mathbb{C}^2\otimes\mathbb{C}^2$ by
  \begin{equation}
    V|j\rangle|k\rangle = \frac12|j\rangle |\Psi^-\rangle |k\rangle,
  \end{equation}
  where $|\Psi^-\rangle \equiv \frac1{\sqrt2}(|01\rangle - |10\rangle)$ is the singlet state. This example may be generalised to the Fuss-Catalan planar algebra \cite{jones_planar_1999}.
\end{exmp}
\begin{exmp}
  Also for the case $n=3$ but with $d=3$ define the map $V\colon\mathbb{C}^3\to\mathbb{C}^3\otimes \mathbb{C}^3$ by
  \begin{equation}\label{eq:4colourtensor}
    \langle jk|V|l\rangle = \begin{cases} 0 \quad \text{if $j=k$, $k=l$, or $j=l$,} \\ 1 \quad \text{otherwise.}\end{cases}
  \end{equation}
  This example comes from the $4$-colour theorem \cite{jones_no-go_2016}.
\end{exmp}
\begin{exmp}
  Another example for $n=4$ and $d=3$ is based on the $3$-qutrit code, given by the $4$-index tensor $T$ defined by
  \begin{equation}
    T|x\rangle|y\rangle = |\text{$2x+y$ mod $3$}\rangle|\text{$x+y$ mod $3$}\rangle.
  \end{equation}
\end{exmp}

There are now several constructions and generalisations of perfect tensors. A partial list includes \cite{goyeneche_absolutely_2015,enriquez_maximally_2016,raissi_constructing_2017,li_invariant_2016,peach_tensor_2017,donnelly_living_2016}.

In this paper, we concentrate on the simplest possible case, where the $3$-leg perfect tensor $V$ is additionally \emph{rotation-invariant}, so that the three conditions for perfectness collapse to only the isometry condition. The isometry defined by (\ref{eq:4colourtensor}) provides a nontrivial example of such a perfect tensor.

\subsection{\texorpdfstring{$\text{AdS}_3$}{AdS3} and a little hyperbolic geometry}

The introductory material described here is adapted from
\cite{brill_black_1999,aminneborg_black_1998,carlip_2_1995,carlip_conformal_2005}.

\subsubsection{The hyperbolic plane}
\begin{figure}
  \centering
  \includegraphics{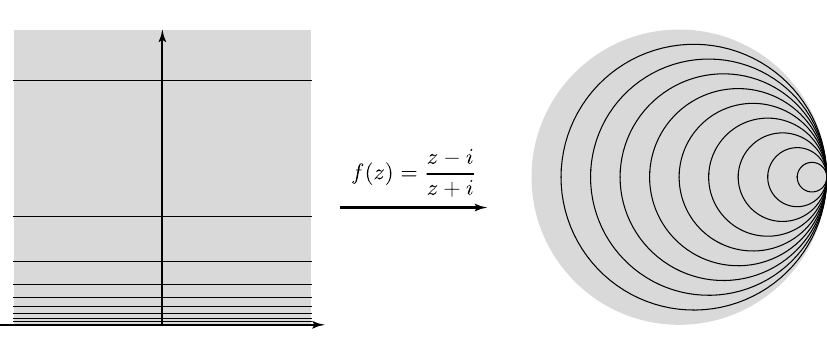}
  \caption{The Cayley transformation is a bijection between the upper-half plane (left) and the Poincar\'e disk (right).}
  \label{fig:cayley}
\end{figure}

Denote the \emph{upper half plane} model of two-dimensional hyperbolic space by $\mathbb{H}^2$ and the \emph{Poincar\'e disc} model of $\mathbb{H}^2$ by $\mathbb{D}$. These two models are related via the conformal transformation
\begin{equation}\label{eq:uhplanetodisc}
	f(z) = \frac{z-i}{z+i},
\end{equation}
known as the \emph{Cayley transformation} (figure \ref{fig:cayley}).

The asymptotic boundary $\partial \mathbb{D}$ of the Poincar\'e disc $\mathbb{D}$ is the circle $S^1$ at infinity. The metric for the disc model is given by
\begin{equation}
	ds^2 = \frac{4dx^2 + 4dy^2}{(1-x^2-y^2)^2}.
\end{equation}
Geodesics in $\mathbb{D}$ are circles which meet the boundary at right angles.

Isometries of the hyperbolic plane are realised by the group $\textsl{PSL}(2,\mathbb{R})$ of M\"obius transformations. These transformations act naturally on the upper half plane model $\mathbb{H}^2$ via \emph{fractional linear transformations}
\begin{equation}
	z\mapsto \frac{az+b}{cz+d}.
\end{equation}
Hence, after conjugation with the conformal map (\ref{eq:uhplanetodisc}), they can also act on $\mathbb{D}$.

\subsubsection{\texorpdfstring{$(2+1)$}{(2+1)}-dimensional anti de Sitter space}
\label{sec:ads3}

Much of the discussion in this paper concerns $(2+1)$-dimensional anti de Sitter space $\text{AdS}_3$. This space may be understood as the quadric surface described by
\begin{equation}
	X^2+Y^2-U^2-V^2=-1
\end{equation}
in flat four-dimensional spacetime with two space and two time dimensions:
\begin{equation}
	ds^2 = dX^2+dY^2-dU^2-dV^2.
\end{equation}
To actually visualise $\mathrm{AdS}_3$ it is convenient to exploit \emph{sausage coordinates} \cite{brill_black_1999,aminneborg_black_1998}
\begin{equation}
	\begin{split}
		X &= \frac{2\rho}{1-\rho^2}\cos(\varphi), \\
		Y &= \frac{2\rho}{1-\rho^2}\sin(\varphi), \\
		U &= \frac{1+\rho^2}{1-\rho^2}\cos(t), \\
		V &= \frac{1+\rho^2}{1-\rho^2}\sin(t),
	\end{split}
\end{equation}
where $0 \le \rho< 1$, $0 \le \varphi < 2\pi$, and $-\pi\le t <\pi$. The metric for $\text{AdS}_3$ in terms of sausage coordinates is given by
\begin{equation}
	ds^2 = -\left(\frac{1+\rho^2}{1-\rho^2}\right)^2 dt^2 + \frac{4}{(1-\rho^2)^2}(d\rho^2 + \rho^2d\phi^2).
\end{equation}
In terms of these coordinates $\text{AdS}_3$ may be visualised as a cylinder whose equal-time slices are copies of the Poincar\'e disc $\mathbb{D}$ and whose end caps are identified.

The boundary of $\text{AdS}_3$ is \emph{timelike} and is seen to be the two-dimensional surface of a cylinder. This is topologically $S^1\times S^1$ and hence identified with the conformal compactification $S^{1,1}\cong S^1\times S^1$ of Minkowski space $\mathbb{R}^{1,1}$. The boundary is called \emph{conformal infinity}.

Geodesics within $\mathrm{AdS}_3$ are found via intersection of the quadric $X^2+Y^2-U^2-V^2=-1$ and hyperplanes containing the origin, i.e., surfaces defined via
\begin{equation}
	a\sin(\alpha)X+a\cos(\alpha)Y-b\sin(\beta)U-b\cos(\beta)V = 0.
\end{equation}
The two-dimensional plane containing a geodesic is thus given by
\begin{equation}
	\frac{2\rho}{1+\rho^2}\sin(\varphi+\alpha) = \frac{b}{a}\sin(t+\beta)
\end{equation}
which is \emph{timelike} if $|b/a| < 1$, \emph{lightlike} if $|b/a|=1$, and \emph{spacelike} if $|b/a|>1$.
Lightlike geodesics in the boundary propagate around the boundary at the speed of light and hence realise spirals or helices.

\subsubsection{Black holes in \texorpdfstring{$2+1$}{2+1} dimensions}\label{sec:blackholes}

Einstein's equations do not admit gravitational wave solutions in $2+1$ dimensions. However, there are black hole type solutions in the case of a negative cosmological constant. These were discovered by Ba\~nados, Teitelboim, and Zanelli \cite{banados_black_1992} (see \cite{carlip_2_1995,carlip_conformal_2005} for further details). The black hole solutions we consider here all arise as quotients of $\mathrm{AdS}_{3}$ by discrete isometries.

The $t=0$ surface of the BTZ black hole solution arises as the result of a quotient by a discrete \emph{hyperbolic} transformation of the Poincar\'e disk: one takes a geodesic $\gamma$, and its image $\gamma'$ under the transformation and identifies them. The BTZ solution arises from the action of the following transformation:
\begin{equation}
	\begin{split}
		X' &= X \\
		Y' &= U\sinh(2\pi\sqrt{M}) + Y\cosh(2\pi\sqrt{M}) \\
		V' &= U\cosh(2\pi\sqrt{M}) + Y\sinh(2\pi\sqrt{M}) \\
		U' &= U.
	\end{split}
\end{equation}

\begin{figure}
  \centering
  \includegraphics{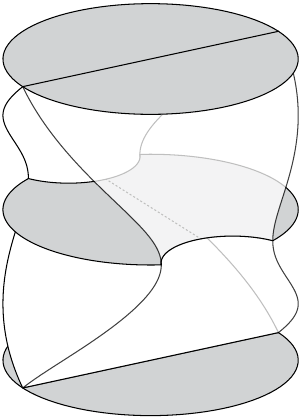}
  \caption{The BTZ black hole in sausage coordinates from $t=-\pi/2$ to $t=\pi/2$. The fundamental domain of the black hole is illustrated as the grey shaded region $t=0$. The two white surfaces represent the development of the two glued geodesics. The solution typically has two event horizons on a constant-time slice. The event horizons coalesce at $t=0$: this is indicated with the dashed line running between the two identified geodesics. This divides the spacetime into two disconnected pieces each with their own boundary at infinity and horizon.}
  \label{fig:btz}
\end{figure}

The topology of the BTZ solution is different from that of $\text{AdS}_3$: the $t=0$ slice is now a cylinder (although it is still locally $\text{AdS}$) and has an event horizon (figure~\ref{fig:btz}).

\subsection{Tessellations}

We build discretised models of $\mathbb{D}$ via \emph{tessellations}, that is, we cover $\mathbb{D}$ with a grid of polygons. Throughout this paper we focus on triangles, but everything we say in the sequel has a natural generalisation to other tessellations via, e.g., pentagons and so on.
\begin{defn}
	A \emph{tessellation} $\mathcal{P}$ of $\mathbb{D}$ (or, a subset $A\subset \mathbb{D}$) is a collection of convex polygons in $\mathbb{D}$ (respectively, $A$) such that
	\begin{enumerate}
		\item the interiors of the polygons in $\mathcal{P}$ are mutually disjoint;
		\item the union of the polygons in $\mathcal{P}$ is $\mathbb{D}$ (respectively, $A$);
		\item the collection $\mathcal{P}$ is locally finite.\footnote{A collection $\mathcal{P}$ of polygons is \emph{locally finite} if and only if for each point $x\in\mathbb{D}$ there is an open neighbourhood of $x$ which nontrivially intersects with only finitely many elements of $\mathcal{P}$.}
	\end{enumerate}
	We say that the tessellation $\mathcal{P}$ is \emph{exact} if and only if every side of a polygon is a side of exactly two polygons in $\mathcal{P}$. A \emph{regular tessellation} of $\mathbb{D}$ is then an exact tessellation consisting of congruent regular polygons.
\end{defn}
We also define a tessellation to be an \emph{ideal regular triangulation} $\tau$ of $\mathbb{D}$ if it is a countable locally finite collection of geodesics in $\mathbb{D}$ such that each connected region in $\mathbb{D}\setminus \tau$ is an \emph{ideal triangle}. (An ideal triangle is a hyperbolic triangle all of whose vertices lie on the boundary of the Poincar\'e disc model.) The \emph{vertices} $\tau^{(0)}$ of the triangulation are the asymptotes of the geodesics comprising the \emph{edges} of the triangulation, regarded as points of the circle at infinity. We denote by $\tau^{(2)}$ the collection of all the complementary triangles in $\mathbb{D}\setminus \tau$.

\begin{figure}
  \centering
  \includegraphics{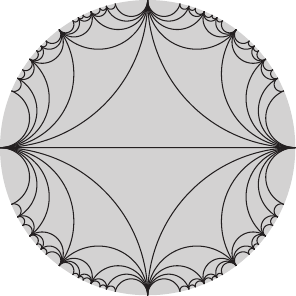}
  \caption{The Farey tessellation.}
  \label{fig:farey}
\end{figure}

\begin{figure}
  \centering
  \includegraphics{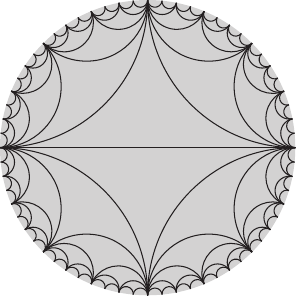}
  \caption{The dyadic tessellation.}
  \label{fig:standarddyadic}
\end{figure}

There are many examples of ideal regular triangulations of $\mathbb{D}$. The \emph{Farey tessellation} $\tau_*$, for instance, is generated by the action of $\mathit{PSL}(2,\mathbb{Z})$ on the basic ideal triangle with vertices at $1,-1,$ and $i$ (figure \ref{fig:farey}).
The Farey triangulation has many mathematical advantages. However, to discuss holographic codes and symmetries it is actually more convenient to use the dyadic tessellation (figure \ref{fig:standarddyadic}).
Its vertices on $\partial \mathbb{D}$ are determined by dyadic subdivision. This means the following: First we realise $\partial \mathbb{D}$ as the unit interval with endpoints identified. Then the coordinates of the points on the boundary have the form $\frac{a}{2^n}$ for $a\in \mathbb{Z}^+$ and $n\in \mathbb{Z}^+$. Both the Farey and dyadic tessellations yield equivalent results in the sequel -- this is a consequence of results of Imbert, see e.g.\ the volume \cite{schneps_geometric_1997} and the papers of Penner, Imbert, and Lochak and Schneps therein. The mapping that identifies the Farey and dyadic tessellations is the Minkowski question mark function $?(x)$.

\begin{figure}
  \centering
  \includegraphics{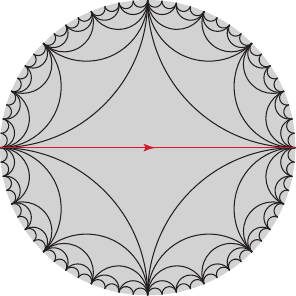}
  \caption{The standard dyadic tessellation with distinguished oriented edge.}
  \label{fig:standarddyadicdoe}
\end{figure}

Just as in Euclidean space, it is often important to distinguish an edge, analogous to the \emph{origin}, to set a reference. This allows us to distinguish transformations preserving the tessellation. In the context of the tessellations considered here this is achieved by choosing an \emph{edge} $e$ along with a preferred \emph{orientation} of $e$ as a \emph{distinguished oriented edge}. The central objects of study in this paper are then \emph{pairs} $(\tau, e)$ of a tessellation $\tau$ together with a specific chosen distinguished oriented edge $e$. The \emph{standard tessellation with distinguished oriented edge} $(\tau_0, e_0)$ is shown in figure~\ref{fig:standarddyadicdoe}.
The edge $e_0$ is taken to be the geodesic joining the points with coordinates $0/1$ and $1/0$ in the upper half plane model.

\begin{figure}
  \centering
  \includegraphics{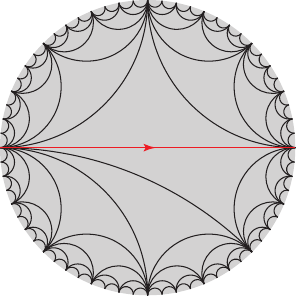}
  \caption{The tessellation we obtain after flipping the geodesic connecting the boundary points $3/4$ and $1$ in the standard dyadic tessellation.}
  \label{fig:standarddyadicdoepachner}
\end{figure}

\begin{figure}
  \centering
  \includegraphics{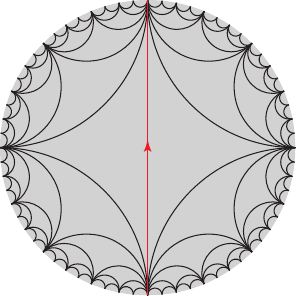}
  \caption{The tessellation we obtain after flipping the distinguished oriented edge.}
  \label{fig:standarddyadicdoepachnerdoe}
\end{figure}

We will also encounter \emph{nonregular} tessellations with distinguished oriented edge. These are all built by applying a \emph{finite} sequence of \emph{Pachner moves} \cite{pachner_p.l._1991} to the standard tessellation. What a Pachner move means in the present context is this: we isolate an ideal quadrilateral formed by two ideal triangles in the tessellation, remove the geodesic joining the diametrically opposed vertices, and then add in a new geodesic between the other pair of opposite vertices. An example of such a tessellation is shown in figure~\ref{fig:standarddyadicdoepachner}.
As long as the diagonal geodesic is \emph{not} the distinguished oriented edge, such a Pachner move is its own inverse. In the case where the diagonal geodesic is the distinguished oriented edge when we apply the Pachner move, we use the orientation given by rotating the edge by $90^{\circ}$. This means that it takes four such Pachner moves to return to the original tessellation (figure~\ref{fig:standarddyadicdoepachnerdoe}).
We call by \emph{admissible} any tessellation $(\tau, e)$ with oriented edge that arises from the standard tessellation via a finite sequence of such Pachner flips.

The main reason for introducing a distinguished oriented edge for a tessellation is that it gives us a mechanism to compare two tessellations without requiring the introduction of a cutoff (a concept we discuss in the next subsection). The way this works is as follows. Suppose that $(\tau, e)$ is an admissible tessellation with oriented edge; we will build a homeomorphism $f\colon S^1\to S^1$ of the boundary $S^1\cong\partial \mathbb{D}$ of the disk which produces this tessellation from the canonical Farey tessellation $(\tau_*,e_*)$. (After you see this construction you will be able to compose two such homeomorphisms to build a map between any two tessellations with oriented edge.) This works because all of the triangles in our tessellations are ideal; therefore they are unambiguously specified once we say where the boundary points are located. Any homeomorphism $f\colon S^1\to S^1$ produces a new (not necessarily admissible) tessellation $(f(\tau_*), f(e_*))$ with oriented edge.

\begin{figure}
  \centering
  \includegraphics{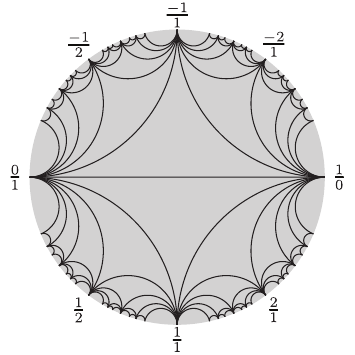}
  \caption{All vertices of the Farey tessellation on the boundary circle can be labelled as follows: start with the three points $0/1$, $1/0$, and $1/1$; then given any pair of labelled vertices $(p/q, r/s)$ of a triangle, label the third vertex in that triangle with the fraction $(p+r)/(q+s)$.}
  \label{fig:LabelledFareyTriangulation}
\end{figure}

The first step in the construction of the homeomorphism $f$ producing $(\tau, e)$ from the Farey tessellation is to label all the vertices of the Farey tessellation with rational numbers according to the following recipe. In the upper half plane model the vertex at $z = 1$ corresponds to $x=\infty$ of $\mathbb{H}^2$, $-1$ to the origin $x=0$ of $\mathbb{H}^2$, and $\sqrt{-1}$ to $x = -1$. We use this identification to iteratively label the vertices of the Farey triangulation: take the edge connecting $0/1$ and $\infty \equiv 1/0$ in the upper half plane model and label the third vertex underneath corresponding to $1/1 = (0+1)/(1+0)$ in the upper half plane. Now continue this process: for every pair of previously labelled vertices $(p/q, r/s)$ of a triangle, label the remaining vertex with the \emph{mediant} $(p+r)/(q+s)$. This process leads to a bijection between the rational numbers $\mathbb{Q}$ and the vertices $\tau_*^{(0)}$, see figure~\ref{fig:LabelledFareyTriangulation}.
We can now recursively build the map $f$: start with the endpoints $\tfrac{0}{1}$ and $\tfrac{1}{0}$: these are mapped to the endpoints of the distinguished oriented edge $e$ in $(\tau, e)$. There is always triangle to the right of the distinguished oriented edge in any tessellation. Let $f$ map the point $\tfrac{1}{1}$ to the vertex of this triangle which isn't on $e$. Do the same with the triangle to the left. Now recursively visit all of the triangles to the left and right, all the while mapping the corresponding vertices from the Farey triangulation to the vertices of the new triangles. This procedure creates a bijection between the vertex sets.
It turns out that this identification may be extended to a homeomorphism $f_{(\tau,e)}$ called the \emph{characteristic mapping} of $(\tau,e)$ \cite{penner_universal_1993,schneps_geometric_1997}. The argument presented here is a powerful manifestation of the bulk boundary correspondence: we've established a bijection between two sets, namely, the \emph{bulk}
\begin{equation*}
	\mathsf{Tess} = \{\text{tessellations of $\mathbb{D}$ with distinguished oriented edge}\},
\end{equation*}
and a set of objects acting on the \emph{boundary}, i.e.,
\begin{equation*}
	\mathrm{Homeo}_+ = \{\text{orientation-preserving homeomorphisms of $\partial \mathbb{D}$}\}.
\end{equation*}
This is the content of the following theorem.
\begin{thm}[Penner \cite{penner_universal_1993,schneps_geometric_1997}]\label{thm:penner}
	The characteristic mapping $(\tau,e) \mapsto f_{(\tau,e)}$ induces a bijection between $\mathsf{Tess}$ and $\mathrm{Homeo}_+$.
\end{thm}

\subsection{Cutoffs}
A crucial role throughout this work is played by \emph{cutoffs}. What are these? We think of imposing a UV cutoff on the system defined on a timeslice $\mathbb{D}$ by truncating it to a region whose boundary is in the interior of $\mathbb{D}$ apart from a finite number of isolated points on the boundary. A consequence of this is that we have restricted the system to have finite volume. Such a cutoff need not be rotation invariant, indeed, it is very convenient to allow the cutoff boundary to have angular dependence.

\begin{figure}
  \centering
  \includegraphics{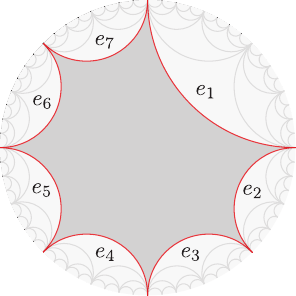}
  \caption{Example of a cutoff: A convex finite area bounded by a closed curve of finitely many geodesics of a given tessellation.}
  \label{fig:PoincareDiscTriangulationCutoffRegion}
\end{figure}

We define a cutoff as follows: Take a list of geodesics $\gamma \equiv (e_1,e_2,\ldots, e_n)$, where the geodesics come from a tessellation $\tau$ of $\mathbb{D}$. Every geodesic partitions the disc into two halfspaces. The \emph{cutoff} $A_\gamma$ associated with $\gamma$ is the finite-volume convex region that is given by the intersection of the halfspaces of the geodesics in $\gamma$. The requirement of finite volume stems from the fact that we want to exclude regions which include a subset of the boundary with nonzero measure. We further require that the geodesics (together with the endpoints on $\partial \mathbb{D}$) comprising the boundary $\partial A_\gamma$ of $A_\gamma$ form a clockwise oriented cycle. (The orientation dictates which side of the geodesic the halfspace is on.) An example of a cutoff with boundary $\gamma = (e_1,e_2,\ldots, e_7)$ is shown in figure~\ref{fig:PoincareDiscTriangulationCutoffRegion}.
For simplicity we'll also refer to the boundary $\gamma$ defining a cutoff $A_\gamma$ as a cutoff.

A important feature of the set of all such cutoffs $\mathcal{P}$ is that it is a \emph{directed set}. This means that we can define a \emph{partial order} $\preceq$ on $\mathcal{P}$ where we say that a cutoff $\gamma$ is \emph{smaller than} the cutoff $\gamma'$, written $\gamma \preceq \gamma'$, if $A_\gamma\subseteq A_{\gamma'}$.
\begin{figure}
  \centering
  \includegraphics{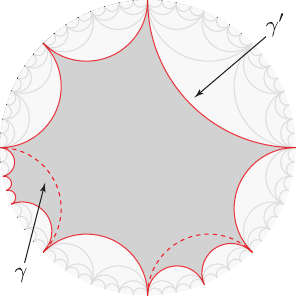}
  \caption{An example of a pair of cutoffs $\gamma\preceq\gamma'$; here $\gamma$ is the original cutoff illustrated in the previous figures, and when it differs from $\gamma'$, it is illustrated with a dashed line.}
  \label{fig:PoincareDiscTriangulationBiggerCutoffRegion}
\end{figure}
Further, given two cutoffs $\gamma$ and $\gamma'$ we can always find a third cutoff $\gamma''$ such that $\gamma \preceq \gamma''$ and $\gamma'\preceq \gamma''$. It is worth noting that since all of our tessellations agree with the dyadic tessellation $\tau_0$ sufficiently close to the boundary (they only differ by a finite number of edge flips/Pachner moves), we can always find a cutoff $\gamma''$ in the standard tessellation $\tau_0$ bigger than $\gamma$ and $\gamma'$ coming from arbitrary tessellations (figure~\ref{fig:PoincareDiscTriangulationBiggerCutoffRegion}).

\subsection{Holographic states}
\begin{figure}
  \centering
  \includegraphics{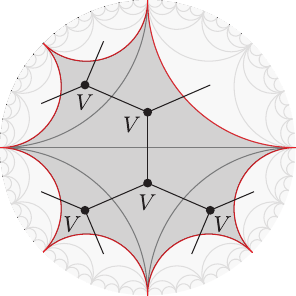}
  \caption{We can form holographic states by placing a fixed perfect tensor $V$ in every triangle of a cutoff coming from a given tessellation, and contracting tensor indices whenever two triangles share a common edge. Since we have taken $V$ to be rotation invariant the orientation of $V$ does not matter.}
  \label{fig:PoincareDiscTriangulationCutoffRegionTNS}
\end{figure}

Given a perfect tensor $V$, a tessellation $\tau$, and a cutoff $\gamma$ coming from $\tau$, we can build a special quantum state $|\psi_\gamma\rangle$ according to the following recipe: for every triangle inside $A_\gamma$ we associate one copy of $V$, with one leg per edge, and for every adjacent pair of triangles we contract the legs of $V$ associated with the common edge, as in figure~\ref{fig:PoincareDiscTriangulationCutoffRegionTNS}.

\begin{figure}
  \centering
  \includegraphics{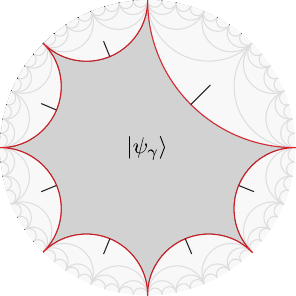}
  \caption{Holographic state.}
  \label{fig:HolographicState}
\end{figure}

We can associate a Hilbert space $\mathcal{H}_\gamma$ with each cutoff $\gamma = (e_1,e_2,\dotsc, e_n)$ in a natural way: Simply take the tensor product of the leg Hilbert space $\mathbb{C}^d$ over each edge of the cutoff $\gamma$, i.e.,
\begin{equation}
	\mathcal{H}_\gamma = \bigotimes_{j=1}^n \mathbb{C}^d.
\end{equation}
If we bipartition this boundary system $\{j_1, j_2, \dotsc, j_n\} \equiv A\cup A^c$ with $A \equiv \emptyset$, then we can regard the tensor network associated to $\tau$ and $\gamma$ as a state $|\psi_\gamma\rangle \in \mathcal{H}_\gamma$, called the \emph{holographic state} (figure~\ref{fig:HolographicState}).
In this way we see that a holographic state associated with a tessellation $\tau$ and a perfect tensor $V$ is really a \emph{family} of states $|\psi_\gamma\rangle$, one per cutoff $\gamma$ from the tessellation $\tau$.

\section{The semicontinuous limit}

A holographic state $|\psi_\gamma\rangle$ built from a perfect tensor $V$ and a tessellation $\tau$ with cutoff $\gamma$ should, in a sense to be specified presently, be \emph{equivalent} to a holographic state $|\psi_{\gamma'}\rangle$ built from the same tessellation and tensor $V$ but with a \emph{larger} cutoff $\gamma'\succeq \gamma$. The question is: How can we compare two states in \emph{different} Hilbert spaces? The answer is to build an \emph{equivalence relation} on the set of all boundary Hilbert spaces.

The physical intuition behind our equivalence relation is the following. A cutoff $\gamma$ induces a (generally nonregular) lattice structure on the boundary space. A larger cutoff $\gamma'\succeq \gamma$ induces a finer lattice structure on the boundary, and some of the cells in the original lattice have been subdivided, or \emph{fine-grained}. In the real-space picture we are working with here, a fine-graining corresponds to a real-space renormalisation group transformation, which is taken to be an \emph{isometry}.

We now have a viable notion of equivalence between two states $|\phi_\gamma\rangle \in \mathcal{H}_\gamma$ and $|\psi_{\gamma'}\rangle \in \mathcal{H}_{\gamma'}$ in two possibly \emph{different} Hilbert spaces. First suppose that $\gamma \preceq \gamma'$. In this case there should be a fine-graining isometry $T^{\gamma}_{\gamma'}\colon\mathcal{H}_\gamma \to \mathcal{H}_{\gamma'}$ which fine-grains any state $|\phi_\gamma\rangle \in \mathcal{H}_\gamma$ into a new state $T^{\gamma}_{\gamma'}|\phi_\gamma\rangle$ living in the finer Hilbert space $\mathcal{H}_{\gamma'}$. Although initially these two states are \emph{mathematically} different, they are \emph{physically equivalent}. From the viewpoint of physics, they represent the \emph{same state} -- just living in Hilbert spaces with different cutoffs. In this context we think of the fine-graining operation $T^{\gamma}_{\gamma'}$ as adding no further information -- i.e., correlations -- above the cutoff $\gamma$. We thus have a method to compare $|\phi_\gamma\rangle$ and $|\psi_{\gamma'}\rangle$: First fine-grain $|\phi_\gamma\rangle$ via $T^{\gamma}_{\gamma'}$ to a state $T^{\gamma}_{\gamma'}|\phi_\gamma\rangle \in \mathcal{H}_{\gamma'}$, then compare $T^{\gamma}_{\gamma'}|\phi_\gamma\rangle$ with $|\psi_{\gamma'}\rangle$ by use of the inner product defined on $\mathcal{H}_{\gamma'}$.

To obtain a general equivalence relation on the set of all Hilbert spaces $\mathcal{H}_\gamma$ associated with cutoffs $\gamma\in \mathcal{P}$ we need to exploit one more feature of the space of cutoffs, namely, that it is a directed set. Suppose we want to compare two states $|\phi_\gamma\rangle \in \mathcal{H}_\gamma$ and $|\psi_{\gamma'}\rangle \in \mathcal{H}_{\gamma'}$ but neither $\gamma \preceq \gamma'$ nor $\gamma'\preceq \gamma$. To do this we realise that there is always a larger cutoff $\gamma''$ which refines both, i.e., $\gamma''\ge\gamma$ and $\gamma''\ge\gamma'$. We then fine-grain both of the states into a common Hilbert space $\mathcal{H}_{\gamma''}$ and compare them there. Their overlap is given by
\begin{equation}
	\langle \phi_\gamma|(T^{\gamma}_{\gamma''})^\dag T^{\gamma'}_{\gamma''}|\psi_{\gamma'}\rangle.
\end{equation}

For all of this to be well defined we need the fine-graining operation $T^{\gamma}_{\gamma'}$ to satisfy two consistency conditions:
\begin{enumerate}
	\item If $\gamma = \gamma'$ then $T^{\gamma}_{\gamma'} = \mathbb{I}$;
	\item For all $\gamma \preceq \gamma'\preceq \gamma''$ we have that
	\begin{equation*}
		T^{\gamma}_{\gamma''} = T^{\gamma'}_{\gamma''}T^{\gamma}_{\gamma'}.
	\end{equation*}
\end{enumerate}
It is straightforward to check that when we have a fine-graining operation obeying conditions (1) and (2) we get a well-behaved equivalence relation $\sim$ on the set $\widehat{\mathcal{H}}$ of all states in some Hilbert space $\mathcal{H}_\gamma$ with some cutoff $\gamma$. The correct space to represent the set of all states with some cutoff is the \emph{disjoint union} of the Hilbert spaces $\mathcal{H}_\gamma$,
\begin{equation}
	\widehat{\mathcal{H}} \equiv \biguplus_{\gamma\in \mathcal{P}} \mathcal{H}_\gamma.
\end{equation}
Why not use the simple union? The problem is when we have two incomparable cutoffs $\gamma$ and $\gamma'$ with the same number of edges: in this case the Hilbert spaces $\mathcal{H}_\gamma$ and $\mathcal{H}_{\gamma'}$ are isomorphic, and hence would collapse to the same element in the standard union. However, these two spaces are physically \emph{very} different; we don't want to forget the cutoff associated to each space. This is achieved by tagging each element of the union with its corresponding cutoff, which is exactly the disjoint union.

When we mod out $\widehat{\mathcal{H}}$ by the equivalence relation $\sim$  we end up with a bona fide Hilbert space
\begin{equation}
	\begin{split}
		\mathcal{H} &= \varinjlim \mathcal{H}_\gamma \\
		&= \left(\biguplus_{\gamma\in \mathcal{P}} \mathcal{H}_\gamma\bigg/\sim\right)^{\| \cdot \|} \\[0.5cm]
		&= \left(\:\rule{0cm}{3em}\text{\begin{minipage}[c][4em][c]{17em}{\footnotesize the disjoint union of $\mathcal{H}_\gamma$ over all $\gamma\in\mathcal{P}$ modulo the equivalence relation that $|\phi_\gamma\rangle \sim |\psi_{\gamma'}\rangle$ if there are $\gamma''\succeq \gamma$ and $\gamma''\succeq\gamma'$ such that $T^{\gamma}_{\gamma''}|\phi_\gamma\rangle=T^{\gamma'}_{\gamma''}|\psi_{\gamma'}\rangle$.}\end{minipage}}\:\right)^{\| \cdot \|}
	\end{split}
\end{equation}
Here $(\dots)^{\| \cdot \|}$ denotes the completion with respect to the standard norm $\| \cdot \|$.
The Hilbert space $\mathcal{H}$ is known as the \emph{direct limit} of the directed system $(\mathcal{H}_\gamma, T^{\gamma}_{\gamma'})$ of Hilbert spaces. It is an infinite-dimensional separable Hilbert space.

\begin{figure}
  \centering
  \includegraphics{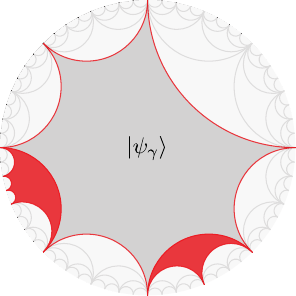}
  \caption{Holographic state on bigger cutoff.}
  \label{fig:HolographicStateBiggerCutoff}
\end{figure}

There is a great deal of arbitrariness in choosing the fine-graining isometries. However, in our case we want the holographic states $|\psi_\gamma\rangle$ to all be equivalent. This is achieved by setting, for $\gamma \preceq \gamma'$, the isometry $T^{\gamma}_{\gamma'}$ to be the tensor network built from $V$ associated with the region bounded by the curves $\gamma$ and $\gamma'$ (figure~\ref{fig:HolographicStateBiggerCutoff}).

When we take for our directed set of possible cutoffs the set $\mathcal{P}_0$ of cutoffs coming from the standard tessellation $\tau_0$ and we use for $T^{\gamma}_{\gamma'}$ the tensor network built from a perfect tensor $V$, we call the resulting direct limit of Hilbert spaces
\begin{equation}
	\mathcal{H} \equiv \varinjlim \mathcal{H}_\gamma
\end{equation}
the \emph{semicontinuous limit} \cite{jones_no-go_2016}. Elements of the semicontinuous limit are \emph{equivalence classes}
\begin{equation}\label{eq:sclinnerproduct}
	[|\phi_\gamma\rangle] \equiv \left\{(\gamma', |\psi_{\gamma'}\rangle) \,\middle|\,  \text{$T^{\gamma'}_{\gamma''}|\psi_{\gamma'}\rangle = T^\gamma_{\gamma''}|\phi_\gamma\rangle$ for some $\gamma''\in\mathcal{P}_0$}\right\}
\end{equation}
 of states coming from boundary systems with a finite cutoff.

\begin{figure}
  \centering
  \includegraphics{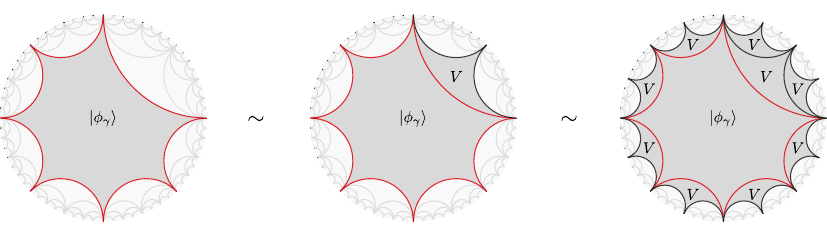}
  \caption{Equivalent state vectors are interpreted as different fine-grainings of the same physical state. The fundamental fine-graining operation is application of the perfect tensor $V$, which acts as an isometry from one qudit to two qudits.}
  \label{fig:PoincareDiscTriangulationEmbeddings}
\end{figure}

What is the physical intuition for a resident of the semicontinuous limit? To get an intuition for this we first note that any vector $|\phi_\gamma\rangle \in\mathcal{H}_\gamma$ has a natural image -- it is \emph{isometrically embedded} -- in $\mathcal{H}$ as
\begin{equation}
	[|\phi_\gamma\rangle].
\end{equation}
You should think of the state $[|\phi_\gamma\rangle] \in \mathcal{H}$ as the \emph{UV completion} of $|\phi_\gamma\rangle$: it is essentially the state $|\phi_\gamma\rangle$ which has been infinitely fine-grained via $T^\gamma_{\gamma'}$ as $\gamma'$ gets closer and closer to the actual boundary $\partial \mathbb{D}$ (figure~\ref{fig:PoincareDiscTriangulationEmbeddings}).

How does one work with the semicontinuous limit $\mathcal{H}$ in practice? Many calculations we need to carry out will not change the cutoff. Since every state $|\phi_\gamma\rangle \in \mathcal{H}_\gamma$ is isometrically embedded in $\mathcal{H}$ we can forget about $\mathcal{H}$ and pretend we are working just in $\mathcal{H}_\gamma$. There are, however, occasions where an operation will lead to a change of cutoff. In this case an initial state $|\phi_\gamma\rangle \in \mathcal{H}_\gamma$ might end up in the space $\mathcal{H}_{\gamma'}$ of boundary states with a different cutoff $\gamma'$. We exploit the inner product defined by (\ref{eq:sclinnerproduct}), i.e.,
\begin{equation}
	\left([|\phi_\gamma\rangle],[|\psi_\gamma\rangle]\right) \equiv \langle \phi_\gamma|(T^{\gamma}_{\gamma''})^\dag T^{\gamma'}_{\gamma''}|\psi_{\gamma'}\rangle,
\end{equation}
and then work with respect to the bounary space $\mathcal{H}_{\gamma''}$. Thus we can effectively work in a finite-dimensional Hilbert space throughout, increasing the cutoff via $T^\gamma_{\gamma'}$ when necessary. Physically this is really no different to how one works with digital images: suppose we have an image defined at some resolution (i.e., cutoff) and we want to prepare an image at a different resolution. Here we fine-grain via \emph{interpolation} -- this is the analogue of $T^\gamma_{\gamma'}$ for digital images -- and then work on a higher-resolution image. This allows us to cut and paste together images with incommensurate resolutions, i.e., \emph{compare} them.

Thus, from now on, we define the semicontinuous limit $\mathcal{H}$ to be the kinematical space for the boundary theory of a holographic state.

\subsection{States with geometry}

Let $(\mathcal{H}_\gamma, T^{\gamma}_{\gamma'})$ be a directed system of Hilbert spaces associated with the holographic state built from a  perfect tensor $V$. Suppose we have a state $[|\phi_\gamma\rangle] \in \mathcal{H}$ in our boundary kinematical space. The state $|\phi_\gamma\rangle$ is allowed to be any possible state in the boundary Hilbert space $\mathcal{H}_\gamma$, which is a $d^{|\gamma|}$-dimensional Hilbert space. However, it could be that $|\phi_\gamma\rangle$ is very special, i.e., it could be that $|\phi_\gamma\rangle$ arises as the contraction of the perfect tensor $V$ according to some tessellation of the region $A_\gamma$, for example:
\begin{equation*}
  \includegraphics[scale=0.8]{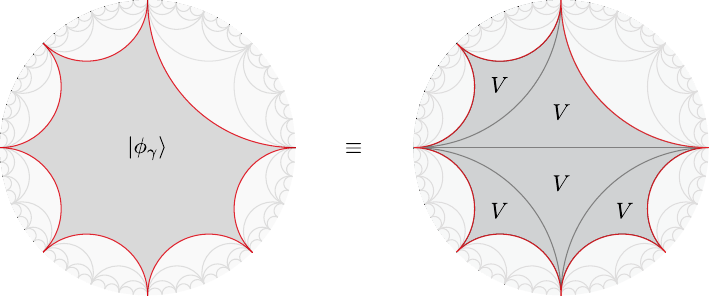}
\end{equation*}
 
Or it could be that $|\phi_\gamma\rangle$ arises from a different tessellation such as
\begin{equation*}
  \includegraphics[scale=0.8]{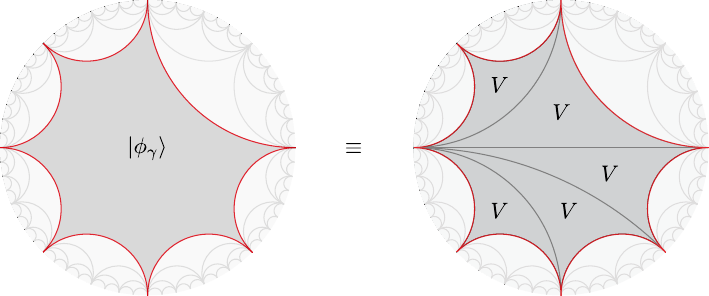}
\end{equation*}

If a state $[|\phi_\gamma\rangle] \in \mathcal{H}$ arises from the contraction of the tensor $V$ according to some tessellation $\tau$ of $A_\gamma$, then we say that it has \emph{geometry $(\tau, A_\gamma)$}.

There is an equivalent way to specify the geometry of a state $[|\phi\rangle] \in \mathcal{H}$ (if it has one) which makes no reference to cutoffs. In this case we say that $[|\phi\rangle]$ has geometry $(\tau, e)$ if it arises as a sequence of states coming from contracting the perfect tensor $V$ according to the tessellation $\tau$. It is not immediate that this approach gives a well-defined state within the semicontinuous limit Hilbert space $\mathcal{H}$. That this is so can be seen as follows. Given an admissible tessellation $(\tau,e)$, we know that far enough away from the origin the tessellation $\tau$ agrees with the standard tessellation $\tau_0$. This is because an admissible tessellation is generated by a finite number of Pachner flips. Now choose a common cutoff $\gamma$ from both $\tau$ and $\tau_0$ (which is guaranteed to exist for sufficiently large $A_\gamma$). Consider the tessellation $\tau$ restricted to $A_\gamma$: this is an instance of the previous scenario, so define the state $|\phi_\gamma\rangle \in \mathcal{H}_\gamma$ by contracting the perfect tensor $V$ according to the tessellation $\tau$ and $A_\gamma$. This recipe is well defined because increasing the cutoff corresponds precisely with the equivalence relation $\sim$ employed to define $\mathcal{H}$ in the first place.

We now introduce a distinguished \emph{subset} $\mathcal{G}$ of $\mathcal{H}$ defined to be the set of all states with some geometry $(\tau, e)$, or equivalently, $(\tau, A_\gamma)$.

\section{Dynamics as a unitary representation of symmetries}

What does it mean for a quantum system to ``have dynamics''? One answer is as follows. The operation ``wait for $t$ units of time'' is a \emph{symmetry} in quantum mechanics, in the sense that waiting $t$ units of time should not change the inner products between pairs, i.e., distinguishability, of states. Hence, thanks to Wigner's theorem (see, e.g., \cite{weinberg_quantum_1996}), this operation must be represented by a (projective) unitary or antiunitary operator $U_t$ on the kinematical Hilbert space $\mathcal{H}$.\footnote{We focus only on the unitary case from now on.} Actually, we get something a little stronger by imposing the condition that first waiting $t_1$ units of time and then waiting $t_2$ units of time is equivalent to waiting $t_1+t_2$ units of time. Further demanding that waiting $0$ units of time corresponds to the identity yields the observation that a quantum system with kinematical Hilbert space $\mathcal{H}$ ``has dynamics'' if it affords a (projective) unitary representation of the \emph{time translation group} $\mathbb{R}$, i.e., we have a family of unitary operators $U_t\colon\mathbb{R}\to \mathcal{U}(\mathcal{H})$ such that
\begin{equation}
	U_{t_1+t_2} = e^{i\phi(t_1,t_2)}U_{t_1}U_{t_2}, \qquad t_1,t_2\in\mathbb{R},
\end{equation}
where $U_0 = \mathbb{I}$ and $\phi$ must satisfy some nontrivial conditions.

It is straightforward to find many such representations for any quantum system: just choose a random Hermitian operator on $\mathcal{H}$ and build
\begin{equation}
	 U_t = e^{-itH}.
\end{equation}
This construction is rather arbitrary and therefore not satisfying. To get a more interesting answer we need to impose additional \emph{constraints} on what we want our dynamics to do. These constraints are typically that the system must exhibit more than just the time-translation symmetry. Indeed, we usually demand that relativistic quantum systems exhibit the full group of Poincar\'e symmetries. Thus, arguing as above, we would say our quantum system is Poincar\'e invariant if we can find a (projective) unitary representation of the (universal cover of the) Poincar\'e group $\mathbb{R}^{1,3}\rtimes\textsl{SL}(2,\mathbb{C})$. This group contains, as a subgroup, our original group $\mathbb{R}$ of time-translation symmetries. However, it is important to note that the group of time-translation symmetries \emph{does not commute} with general Poincar\'e transformations. Thus it is not sufficient to find a Hamiltonian $H$ that commutes with the generators of boosts etc. This is precisely why building representations of the Poincar\'e group is much harder than building symmetric models in nonrelativistic quantum mechanics: we have to do everything at once in the relativistic setting.

In this paper we are looking for something much stronger than just a relativistic quantum system. Indeed, we want our quantum system to correspond to a $(1+1)$-dimensional \emph{conformally invariant} quantum system. A rather strong interpretation of this is that our quantum system should not only give us a unitary representation of the Poincar\'e group for $\mathbb{R}^{1,1}$ but also a unitary representation of local conformal transformations. The group $\conf(\mathbb{R}^{1,1})$ of all local conformal transformations of $\mathbb{R}^{1,1}$ is given by (see, e.g., \cite{schottenloher_mathematical_2008})
\begin{equation}
	\conf(\mathbb{R}^{1,1}) = \left(\diff_+(\mathbb{R})\times \diff_+(\mathbb{R})\right)\cup \left(\diff_-(\mathbb{R})\times \diff_-(\mathbb{R})\right)
\end{equation}
Hence the group of \emph{all} conformal diffeomorphisms consists of two connected components. We now restrict our attention to the connected component $\diff_+(\mathbb{R})\times \diff_+(\mathbb{R})$. This group may be understood in terms of $\diff_+(\mathbb{R})$, which is concrete enough. However, $\diff_+(\mathbb{R})$ is extremely large and we rather study the conformal diffeomorphisms of the conformal compactification of $\mathbb{R}^{1,1}$. Thus the group of orientation-preserving conformal diffeomorphisms of the conformal compactification $S^{1,1}$ is isomorphic to the group
\begin{equation}
	\left(\diff_+(S^1)\times \diff_+(S^1)\right)\cup \left(\diff_-(S^1)\times \diff_-(S^1)\right).
\end{equation}
One then \emph{redefines} ``the'' conformal group $\conf(\mathbb{R}^{1,1})$ of $\mathbb{R}^{1,1}$ to be the connected component of $\diff(S^{1,1})$, namely
\begin{equation}
	\conf(\mathbb{R}^{1,1}) \cong \diff_+(S^1)\times \diff_+(S^1).
\end{equation}
Unitary representations of $\conf(\mathbb{R}^{1,1})$ are all built from representations of $\diff_+(S^1)$, known as the \emph{chiral conformal group}. So, as is typical, we focus on $\diff_+(S^1)$.

Thus our goal, at its most ambitious, is to find a (projective) unitary representation of $\diff_+(S^1)$ on our kinematical semicontinuous limit space $\mathcal{H}$. If we could actually do this then we'd have arguably built a full conformal field theory: to get the Hamiltonian, for example, one just needs to differentiate the representation of the one-parameter group of time translations.

That the semicontinuous limit Hilbert space $\mathcal{H}$ might be a natural place to look for a unitary representation of $\diff_+(S^1)$ comes from the observation that $\diff_+(S^1)\subset \mathrm{Homeo}_+$ which, in turn, may be identified with the space $\mathsf{Tess}$. This leads to the naive idea of representing the action of $f\in\diff_+(S^1)$ on $\mathcal{H}$ by an operator $\pi(f)$ that takes a state $|\psi_{(\tau,e)}\rangle$ with geometry $(\tau, e)$ to a state with geometry $(f(t),f(e))$ and then extending by linearity. This idea \emph{almost works}, but runs into the problem that when $f\in\diff_+(S^1)$ acts on the boundary $S^1$ it typically takes an admissible tessellation with distinguished oriented edge $(\tau, e)$ to an \emph{inadmissible} tessellation. There seems no easy way to add in states with the geometry of these inadmissible tessellations without leading to a nonseparable Hilbert space!

The approach we take instead is to understand exactly what transformations \emph{do} preserve the admissability of a tessellation and instead focus on the group these transformations generate. Astonishingly, this group, known as \emph{Thompson's group $T$}, while not containing $\diff_+(S^1)$, does contain sequences which can approximate any $f\in\diff_+(S^1)$ arbitrarily well.

\subsection{Thompson's group \texorpdfstring{$T$}{T}}

We review in this subsection the definition and some of the basic properties of Richard Thompson's groups $F$ and $T$. The material presented here is adapted from the canonical reference \cite{cannon_introductory_1996} of Cannon, Floyd, and Parry.

We start with the definition of a group known as Thompson's group $F$, which is a subgroup of the group $T$ which we work with in the sequel.
\begin{defn}
	We call by \emph{Thompson's group $F$} the group of piecewise linear homeomorphisms from $[0,1]$ to itself which are differentiable except at finitely many dyadic\footnote{A \emph{dyadic rational} is a rational number of the form $a/2^n$, with $a$ and $n$ integers.} rational numbers and such that on the differentiable intervals the derivatives are powers of $2$.
\end{defn}

\begin{rem}
	That $F$ is indeed a group follows from the following observations. Let $f\in F$. Since the derivative of $f$, where it is defined, is always positive, it preserves the orientation of $[0,1]$. Suppose that $0 = x_0 < x_1 < \cdots < x_n = 1$ are the points where $f$ is not differentiable. Then
	\begin{equation}
		f(x) = \begin{cases}
			a_1x, \quad & x_0\le x \le x_1, \\
			a_2x+ b_2, \quad & x_1\le x \le x_2, \\
			& \quad\quad\vdots  \\
			a_nx+ b_n, \quad & x_{n-1}\le x \le x_n,
		\end{cases}
	\end{equation}
	where, for all $j=1, 2, \ldots, n$, $a_j$ is a power of two and $b_j$ is a dyadic rational (and we set $b_1 = 0$). The inverse $f^{-1}$ also has power-of-two derivatives except on dyadic rational points and, since $f$ maps the set of dyadic rationals to itself, we deduce that $F$ is a group under composition.
\end{rem}

To define $T$ we regard $S^1$ as the interval $[0,1]$ with the endpoint $1$ identified with $0$. Thompson's group $T$ is then the collection of piecewise linear homeomorphisms from $S^1$ to $S^1$ taking dyadic rational numbers to dyadic rational numbers and which are differentiable except at a possibly finite number of locations which are also dyadic rational, and whose slopes are given by powers of $2$. That $T$ is a group follows from an argument identical to that presented for $F$ above. Since every element of $F$ also satisfies the conditions to be an element of $T$ we have that it is a subgroup $F\le T$.

\begin{figure}
  \centering
  \includegraphics{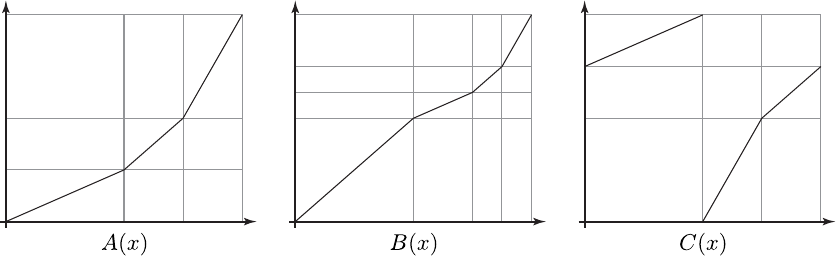}
  \caption{The three generators of Thompson's group $T$. $A$ and $B$ together generate Thompson's group $F$.}
  \label{fig:ThompsonGenerators}
\end{figure}

It turns out \cite{cannon_introductory_1996} that Thompson's group $F$ is generated by two elements, $A$ and $B$, defined by
\begin{equation}
	\begin{split}
		A(x) &=
		\begin{cases}
			\frac12 x, &\quad x\in [0,\tfrac12),\\
			x-\frac14, &\quad x\in [\tfrac12, \tfrac34),\\
			2x-1, &\quad x\in [\tfrac34, 1],
		\end{cases}\\
		B(x) &=
		\begin{cases}
			x, &\quad x\in [0,\tfrac12),\\
			\frac{x}{2}+\frac14, &\quad x\in [\tfrac12, \tfrac34),\\
			x-\frac18, &\quad x\in [\tfrac34, \tfrac78
			), \\
			2x-1, &\quad x\in [\tfrac78, 1].
		\end{cases}
	\end{split}
\end{equation}
Thompson's group $T$ is generated by $A$ and $B$ together with a third element $C$, defined by
\begin{equation}
		C(x) =
		\begin{cases}
			\frac{x}{2} + \frac34, &\quad x\in [0,\tfrac12),\\
			2x-1, &\quad x\in [\tfrac12, \tfrac34),\\
			x-\frac14, &\quad x\in [\tfrac34, 1],
		\end{cases}
\end{equation}
$A$, $B$, and $C$ are illustrated in figure~\ref{fig:ThompsonGenerators}.

\begin{figure}
  \centering
  \includegraphics{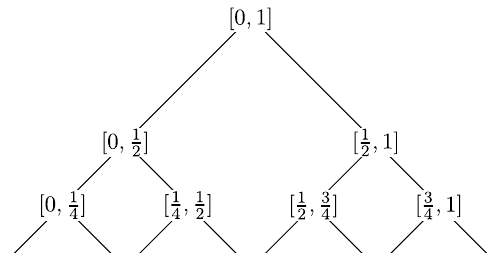}
  \caption{The infinite tree $\mathcal{T}$ of standard dyadic intervals. The two children of every vertex represent the two intervals obtained by dividing the interval corresponding to that vertex in two halves.}
  \label{fig:dyadictree}
\end{figure}

There are many alternative representations that have been developed to work with elements of $F$ and $T$. One of the most convenient for us will be via \emph{tree diagrams}. To describe these we first introduce the tree $\mathcal{T}$ of standard dyadic intervals.
\begin{defn}
	A \emph{standard dyadic interval} is an interval in $[0,1]$ of the form $[\frac{a}{2^n}, \frac{a+1}{2^n}]$, where $a,n\in \mathbb{Z}^{+}$.
\end{defn}
We build the tree $\mathcal{T}$ of standard dyadic intervals by introducing a node for each standard dyadic interval and connecting two nodes if one is included in the other (figure~\ref{fig:dyadictree}). Here one can think of edges as denoting \emph{subdivisions}.
A finite ordered rooted binary subtree with root $[0,1]$ of $\mathcal{T}$ is called a \emph{$\mathcal{T}$-tree}. From now on we suppress the labellings of the nodes via standard dyadic intervals as they may be reconstructed from context. Here are two examples of $\mathcal{T}$-trees:
\begin{equation}
  \includegraphics[scale=0.5]{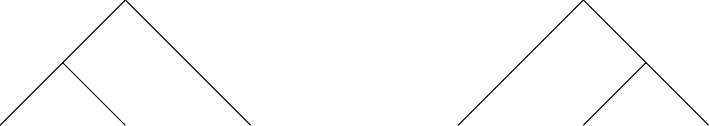}
\end{equation}
There is a one-to-one correspondence between $\mathcal{T}$-trees and certain partitions of the unit interval:
\begin{defn}
	Let $\Gamma \equiv \{[x_0,x_1), [x_1,x_2), \ldots, [x_{n-1},x_{n}]\}$, with $0 = x_0 < x_1 < \cdots < x_n = 1$, be a partition of $[0,1]$. We define $[x_j,x_{j+1}]$, $j = 0,1, \ldots, n-1$, to be the \emph{intervals} of the partition (note the inclusion of the end points). A partition $\Gamma$ of $[0,1]$ is called a \emph{standard dyadic partition} if all the intervals of $\Gamma$ are standard dyadic intervals.
\end{defn}
The leaves of a $\mathcal{T}$-tree describe the intervals of a standard dyadic partition and there is a bijection between standard dyadic partitions and $\mathcal{T}$-trees.

The utility of $\mathcal{T}$-trees in discussing Thompson's groups comes from the following lemma.
\begin{lem}\label{lem:stdyadic}
	Let $f\in F$. There is a standard dyadic partition $\Gamma = \{0 = x_0 < x_1 < \cdots < x_n = 1\}$ such that $f$ is linear and differentiable on every interval of $\Gamma$ and $f(\Gamma) = \{f(0) = f(x_0) < f(x_1) < \cdots < f(x_n) = f(1)\}$ is also a standard dyadic partition.
\end{lem}
This lemma provides us with the motivation to introduce the notion of a \emph{tree diagram}. This is a pair $(R,S)$ of $\mathcal{T}$-trees such that $R$ and $S$ have the same number of leaves. You should think of this pair as a fraction $\frac{R}{S}$. The tree $R$ is the \emph{domain tree} or \emph{numerator tree} and $S$ is the \emph{range tree} or \emph{denominator tree}. Here is an example of a tree diagram:
\begin{equation}
	\left(\raisebox{-2pt}{\includegraphics{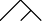}, \includegraphics{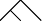}}\right).
\end{equation}
Because of lemma~\ref{lem:stdyadic} we know that there exist standard dyadic partitions $\Gamma$ and $f(\Gamma)$ such that $f$ is differentiable and linear on the intervals of $\Gamma$ and maps them to the intervals of $f(\Gamma)$. We associate to $f$ the tree diagram $(R,S)$ where we get $R$ from the partition $\Gamma$ by representing it as a $\mathcal{T}$-tree, and $S$ is similarly the $\mathcal{T}$-tree associated to $f(\Gamma)$.

Note that one can associate many different tree diagrams $(R,S)$ to $f\in F$: these can be obtained by simultaneously adjoining \emph{carets} to the leaves of $R$ and $S$. What this means is that we take the $j$th leaves of both $R$ and $S$ and simultaneously adjoin the binary tree with two leaves to these leaves. This process is illustrated here, highlighting the adjoined carets in red:
\begin{equation}
	\left(\raisebox{-2pt}{\includegraphics{leftATree.pdf}, \includegraphics{rightATree.pdf}}\right)\mapsto \left(\raisebox{-4pt}{\includegraphics{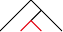}, \includegraphics{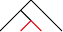}}\right).
\end{equation}
The caret adjunction process builds a new standard dyadic partition $\Gamma'$ from $\Gamma$ which has the same intervals as $\Gamma$ except that the $j$th interval has been symmetrically subdivided into two intervals. Call the interval in $\Gamma$ represented by the $j$th leaf $I$ and the corresponding interval in $f(\Gamma)$ by $J=f(I)$. The new intervals in the partition $\Gamma'$ represented by the leaves of the adjoined carets are called $I_1$ and $I_2$. Since the map $f$ is linear and differentiable on the interval $I$ represented by the $j$th leaf, we deduce that $f(I_1) = J_1$ and $f(I_2) = J_2$. Thus the tree diagram $(R',S')$ arising from the caret adjunction is also a tree diagram for $f$.

One can \emph{reduce} a tree diagram $(R,S)$ by eliminating common carets. If there are no common carets that can be eliminated, the diagram is said to be \emph{reduced}. It turns out \cite{cannon_introductory_1996} that there is exactly one reduced tree diagram for every $f\in F$ and vice versa.

Here are the reduced tree diagrams for $A$ and $B$:
\begin{equation}
	A = \left(\raisebox{-2pt}{\includegraphics{leftATree.pdf}, \includegraphics{rightATree.pdf}}\right), \quad B= \left(\raisebox{-4pt}{\includegraphics{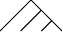}, \includegraphics{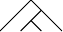}}\right).
\end{equation}

This entire discussion can be repeated for the case of Thompson's group $T$ with essentially no modification. The only difference is due to the fact that elements of $T$ can move the origin. To keep track of this, when an element of $T$ \emph{does} map the origin to a different point we denote the image of the interval in the domain tree in the range tree with a small circle. This is illustrated for $C$ here:
\begin{equation}
	C = \left(\raisebox{-2pt}{\includegraphics{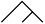}, \includegraphics{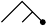}}\right).
\end{equation}

Since we are free to adjoin as many carets as we like, we can understand Thompson's groups $F$ and $T$ as \emph{rewriting rules} for infinite trees: an element of one of Thompson's groups will take an infinitely large domain tree, cut off the infinite base tree leaving a finite domain tree, replace the reduced domain tree with the range tree, and then add the infinite base by adjunction of carets. This action is in sharp contrast to the way other natural groups act on trees, in particular, the $p$-adic groups which have featured in a related model of the AdS/CFT correspondence \cite{gubser_p-adic_2017,heydeman_tensor_2016}.

\subsection{Approximating diffeomorphisms}

In this subsection we describe a fundamental result which explains the precise connection between Thompson's group $T$ and the chiral conformal group. Both groups are subsets of the group $\mathrm{Homeo}(S^1)$ of homeomorphisms of the circle, and our ambition is that by studying $T$ we can learn about $\diff_+(S^1)$ because the two groups are morally ``close'' to one another (in the extremely coarse sense that they are subgroups of the same group).

The most direct hope for using Thompson's group $T$ to study the chiral conformal group $\diff_+(S^1)$ would have been, e.g., that $T$ is a subgroup of the conformal group $T\le \diff_+(S^1)$. However, this hope was doomed to failure from the outset as the elements of $T$ are not differentiable. The next best thing, therefore, is to try and approximate one group by the other. This is indeed possible and is captured by the following striking proposition \cite{BieriStrebel16,stiegemann2018}.

\begin{prop}
	Let $f\in\diff_+(S^1)$. Then there exists a sequence $g_n\in T$, $n\in \mathbb{N}$, such that
	\begin{equation}
		\lim_{n\to \infty}\|f-g_n\|_\infty = \lim_{n\to \infty}\sup_{x\in S^1} |f(x)-g_n(x)| = 0.
	\end{equation}
\end{prop}

Thanks to this proposition we can be encouraged that studying $T$ will give us some insight into $\diff_{+}(S^1)$. Suppose we have a general purpose way to build (projective) unitary representations $\pi$ of $T$, then, supposing that the representation $\pi$ is ``sufficiently continuous'' then we'd have a general purpose procedure to build representations of $\diff_{+}(S^1)$ according to
\begin{equation}
	\pi(f)\, \text{``$\equiv$''}\, \lim_{n\to \infty} \pi(g_n).
\end{equation}
The scare quotes here indicate that we currently do not understand how to find such sufficiently continuous representations. If we could do this, however, then we'd have constructed a new procedure to build conformal field theories.

The emphasis in this paper is to promote a bug to a feature and study projective unitary representations of $T$ in their own right as a toy model of the AdS/CFT correspondence. The benefit of this is the resulting representations are so explicit that we can calculate everything of interest explicitly.

\subsection{The action of Thompson's group \texorpdfstring{$T$}{T} on tessellations}

\begin{figure}
  \centering
  \includegraphics{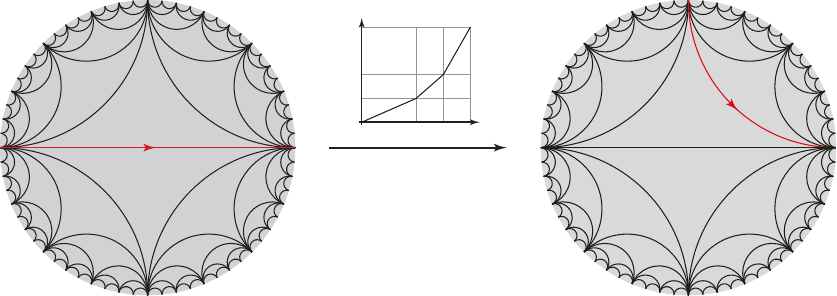}

  \vspace*{1cm}

  \includegraphics{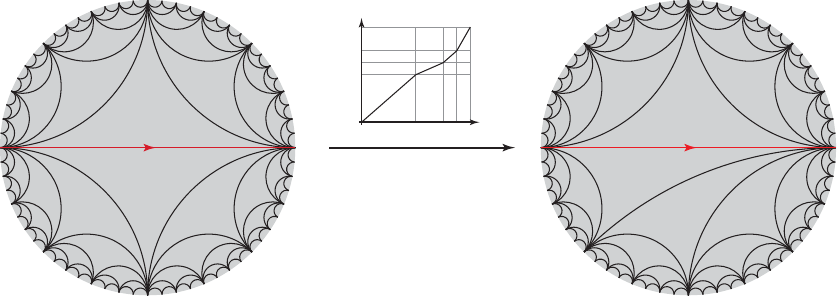}

  \vspace*{1cm}

  \includegraphics{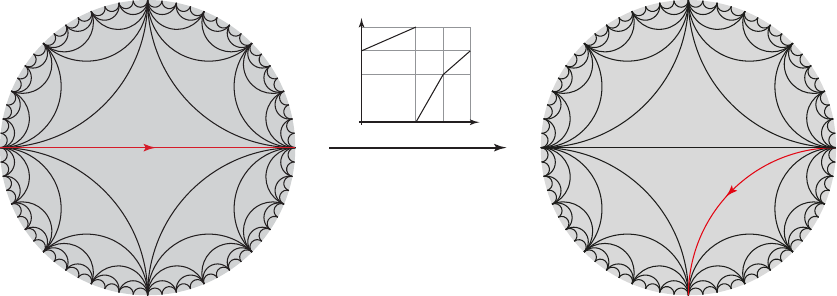}
  \caption{The action of the three generators $A$, $B$, and $C$ of Thompson's group $T$ on the standard dyadic tessellation.}
  \label{fig:ThompsonAction}
\end{figure}

Thompson's group $T$ acts in a natural way on the boundary $S^1 = \partial\mathbb{D}$ of the Poincar\'e disc and, in particular, on points with dyadic rational coordinates. According to this action our standard dyadic tessellation with distinguished oriented edge $(\tau,e)$ is mapped to another tessellation with vertices on dyadic rationals. It is a remarkable consequence of the work of Imbert, Penner, and Lochak and Schneps \cite{schneps_geometric_1997} that every element of $T$ takes any admissible tessellation with distinguished oriented edge to another admissible tessellation with distinguished oriented edge, i.e., via Pachner flip. We don't revisit this proof here. Instead, we present the action of the generators $A$, $B$, and $C$ on the standard dyadic tessellation as this is illustrative enough to imagine how an arbitrary Thompson group element acts (figure~\ref{fig:ThompsonAction}).

\subsection{Unitary representations of \texorpdfstring{$T$}{T} from perfect tensors}

In this subsection we review a general purpose procedure, due to Jones \cite{jones_unitary_2014}, to build unitary representations of $T$ from perfect tensors. There are a couple of ways to understand this construction, by far the most elegant of which is a categorical argument based on the localisation functor \cite{jones_no-go_2016}. A general mathematical framework for this is described in \cite{stiegemann2019}.

The Hilbert space that furnishes our representation is none other than the semicontinuous limit space $\mathcal{H}$. Recall that elements of this space are (equivalence classes of) cutoff+state pairs $(\gamma, |\psi_\gamma\rangle)$, where $|\psi_\gamma\rangle \in \mathcal{H}_\gamma$, subject to the equivalence relation $\sim$. Since our cutoffs $\gamma$ are defined by the dyadic tessellation $\tau$ the endpoints of each geodesic in the cutoff lies on \emph{dyadic rationals}. Thompson's group $T$ acts naturally on geodesics with dyadic rational endpoints by simply moving the endpoints. Hence $T$ acts naturally on cutoffs $\gamma = (e_1, e_2, \ldots, e_n)$ as we simply define $f(\gamma)$ to be the cutoff $(f(e_1), f(e_2), \ldots, f(e_n))$. In this fashion we can allow $T$ to act on elements of $\mathcal{H}$: suppose we have an element of $\mathcal{H}$ with representative $(\gamma, |\psi_\gamma\rangle)$. Then we could build an action by setting
\begin{equation}
	\pi(f)(\gamma, |\psi_\gamma\rangle) \,\text{``$=$''}\,(f(\gamma), |\psi_\gamma\rangle).
\end{equation}
The reason there are scare quotes is due to the fact that since $T$ acts via Pachner flips of the standard dyadic tessellation we are only guaranteed that $f(\gamma)$ is a cutoff coming from an \emph{admissible tessellation}. The only subtlety is that sometimes $f(\gamma)$ will not be a cutoff in $\mathcal{P}_0$, i.e., one coming from the standard dyadic tessellation $\tau_0$. This is a problem for defining an action of $T$ which maps elements of $\mathcal{H}$ to elements of $\mathcal{H}$, because $(f(\gamma), |\psi_\gamma\rangle)$ is a member of $\mathcal{H}$ only when $f(\gamma)\in \mathcal{P}_0$. However, it is always possible to remedy this problem: just take a finer cutoff $\gamma'\succeq \gamma$ such that $f(\gamma')$ also comes from the standard dyadic partition $\tau_0$. That this is always possible follows from Lemma~\ref{lem:stdyadic}: One must first identify cutoffs with \emph{partitions} of the circle $S^1$ via the intervals determined by the endpoints of the geodesics. For example, the cutoff defined by $\gamma = (e_1,e_2,\ldots, e_7)$,
\begin{equation}
\includegraphics[width=40mm]{PoincareDiscTriangulationCutoffRegion.pdf}
\end{equation}
is naturally associated to the partition
\begin{equation}
	\Gamma = \{[0,\tfrac14), [\tfrac14, \tfrac38), [\tfrac38,\tfrac12),[\tfrac12,\tfrac58),[\tfrac58,\tfrac34),[\tfrac34,\tfrac78), [\tfrac78, 1)\}
\end{equation}
of $S^1$.

Therefore, the action of $T$ is defined as follows: first take a representative $(\gamma, |\psi_\gamma\rangle)$ of the element $[|\psi_\gamma\rangle]\in\mathcal{H}$ you want to act on, then refine it to an equivalent representative
\begin{equation}
	(\gamma, |\psi_\gamma\rangle) \sim (\gamma', T^{\gamma}_{\gamma'}|\psi_\gamma\rangle)
\end{equation}
such that $f(\gamma') \in\mathcal{P}_0$, i.e.\ it corresponds to a standard dyadic partition, and then define
\begin{equation}
	\pi(f)([|\psi_\gamma\rangle]) \equiv \pi(f)(\gamma, |\psi_\gamma\rangle) \equiv (f(\gamma'), T^{\gamma}_{\gamma'}|\psi_\gamma\rangle).
\end{equation}
One still needs to check that everything is well defined, i.e., that refinement plays nicely with the action of $T$. This can be done and is detailed in \cite{jones_unitary_2014}.

It is worth working through one example in detail in order to convince yourself that the action $T$ we've defined actually leads to something nontrivial. To do this we focus on the simplest possible example state, namely the state determined by the perfect tensor $V$ itself on an ideal triangle:
\begin{equation*}
  \includegraphics[scale=0.9]{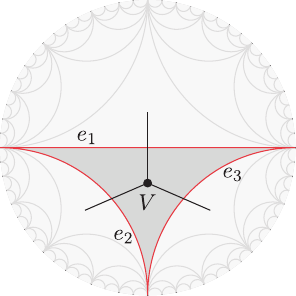}
\end{equation*}
A representative for this state is $(\gamma, |\phi_\gamma\rangle)$ where $\gamma = (e_1, e_2, e_3)$ corresponds to the partition $\{[0,\tfrac12), [\tfrac12, \tfrac34), [\tfrac34, 1)\}$ and $|\phi_\gamma\rangle \in \mathbb{C}^d\otimes\mathbb{C}^d\otimes\mathbb{C}^d$ (note that $\langle jkl|\phi_\gamma\rangle = \frac{1}{\sqrt{d}}{V^j}_{kl}$, i.e., $|\phi_\gamma\rangle$ is proportional to $V$ in order it be a normalised state). We apply the generator $B$ to this representative. The first difficulty we encounter is that $B(\gamma)$ is \emph{not} an admissible region. Therefore, to correctly apply the transformation we first need to refine $\gamma$ via $T^\gamma_{\gamma'} = \mathbb{I} \otimes \mathbb{I}\otimes V$ to a new cutoff
$\gamma' = (e_1, e_2, f_3, f_4)$
corresponding to the partition
$\{[0,\tfrac12), [\tfrac12, \tfrac34), [\tfrac34, \tfrac78), [\tfrac78, 1)\}$,
giving us the equivalent representative
$(\gamma', \mathbb{I}\otimes \mathbb{I}\otimes V|\phi_\gamma\rangle)$
which is illustrated here:
\begin{equation*}
  \includegraphics[scale=0.9]{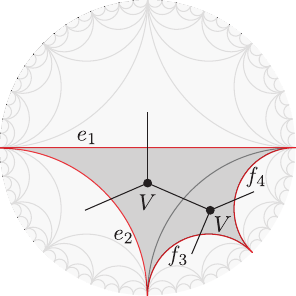}
\end{equation*}

The generator $B$, when applied to $\gamma'$, now does give us an admissible region $B(\gamma')$ corresponding to the new partition $\{[0,\tfrac12), [\tfrac12, \tfrac58), [\tfrac58, \tfrac34), [\tfrac34, 1)\}$. The action of $\pi(B)$ is therefore
\begin{equation}
	\pi(B)(\gamma', \mathbb{I}\otimes \mathbb{I}\otimes V|\phi_\gamma\rangle) = (B(\gamma'), \mathbb{I}\otimes \mathbb{I}\otimes V|\phi_\gamma\rangle).
\end{equation}
The result of the transformation is
\begin{equation*}
  \includegraphics[scale=0.9]{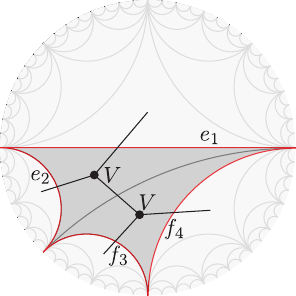}
\end{equation*}

Although it might look as though $B$ did nothing, the pair $(B(\gamma'), \mathbb{I}\otimes \mathbb{I}\otimes V|\phi_\gamma\rangle)$ is in general the representative for a \emph{different state} in the semicontinuous limit space. This is because $B$ moved the legs around: the pair $(B(\gamma'), \mathbb{I}\otimes \mathbb{I}\otimes V|\phi_\gamma\rangle)$ has the same state $\mathbb{I}\otimes \mathbb{I}\otimes V|\phi_\gamma\rangle$ but now associated to a \emph{different region} with boundary $B(\gamma')$. That is, the state $\mathbb{I}\otimes \mathbb{I}\otimes V|\phi_\gamma\rangle$ has been \emph{moved} from the subspace $\mathcal{H}_{\gamma'}$ into a different subspace of $\mathcal{H}$, namely, $\mathcal{H}_{B(\gamma')}$.

Let's see if the new state $\pi(B)(\gamma', \mathbb{I}\otimes \mathbb{I}\otimes V|\phi_\gamma\rangle)$ after the action of $B$ is any different to the original state $(\gamma, |\phi_\gamma\rangle)$: to do this we need to compare them within the same subspace of $\mathcal{H}$. This is rather simple in this case because $\gamma \preceq B(\gamma')$, so all we need to do is refine our original representative into $B(\gamma')$ via $T^{\gamma}_{B(\gamma')} = \mathbb{I}\otimes V\otimes \mathbb{I}$. On this common subspace $\mathcal{H}_{B(\gamma')}$ we find that the original state and the final state have representatives
\begin{equation}
	(B(\gamma'), \mathbb{I}\otimes V\otimes \mathbb{I}|\phi_\gamma\rangle), \quad \text{and} \quad (B(\gamma'), \mathbb{I}\otimes \mathbb{I}\otimes V|\phi_\gamma\rangle),
\end{equation}
respectively. The inner product between the two states is thus
\begin{equation}\label{eq:innerproductrep}
	\langle \phi_\gamma|(\mathbb{I}\otimes V^\dag\otimes \mathbb{I})(\mathbb{I}\otimes \mathbb{I}\otimes V)|\phi_\gamma\rangle
\end{equation}
which, depending on $V$, is not always equal to $1$. In particular, for the example (\ref{eq:4colourtensor}) we find that the inner product is given by $1/2$. Thus we have shown that the action of $T$ on the Hilbert space $\mathcal{H}$ can be nontrivial.

Now that we have an action of $T$ on $\mathcal{H}$ we need to establish that it is unitary. This is relatively easy to do by confirming the invariance of the inner product for all representatives $(\gamma, |\phi_\gamma\rangle)$:
\begin{equation}
	(\pi(f)[|\phi_\gamma\rangle], \pi(f)[|\phi_\gamma\rangle]) = ([|\phi_\gamma\rangle], [|\phi_\gamma\rangle]).
\end{equation}
That this is true is a consequence of the fact that $V$ is an isometry, which in turn follows from the perfect tensor condition. We will not deny the reader the pleasure of confirming this result for themselves.

One might wonder if the unitary representation of $T$ thus constructed is generated by a Hamiltonian. To this end, note that in the continuum case, time evolution amounts to a (strongly continuous) unitary representation $t\mapsto U_t$, $t\in\mathbb{R}$, of the time-translation group $\mathbb{R}$. The Hamiltonian can then be obtained as $H=i\hbar\frac{d}{dt}U_t$, so that $U_t=e^{-iHt/\hbar}$. But what is the analogue of time in our discrete setting? Since the modular group $\mathit{PSL}_2(\mathbb{Z})$ is a subgroup of $T$ (generated by the elements $A$ and $C$ of $T$), the only translations of the form $z\mapsto \frac{az+b}{cz+d}$ with $a,b,c,d\in\mathbb{Z}$ are those for which $a=d=1$ and $c=0$, that is, only translations $z\mapsto z+b$ by integers $b$ are possible. It is therefore reasonable to say that if there is an analogue of time-translations in the discrete theory, it is given by $\mathbb{Z}$. But then the time-derivative $\frac{d}{dt}U_t$ and therefore the notion of Hamiltonian are not well-defined, so we will have to be content with the unitary representation as dynamics. Furthermore, note that the absence of a Hamiltonian is a commonly accepted and expected behaviour of discrete theories such as those of quantum walks or quantum cellular automata, and therefore not at all surprising.

\subsection{The bulk Hilbert space}

In this subsection we build a subspace $\mathcal{H}_{\text{bulk}}\subset \mathcal{H}$ which we later argue corresponds to the Hilbert space for the dual bulk gravitational theory. For the mathematically inclined: what we will do here is build the GNS representation of the von Neumann group algebra built on $T$.

In the previous subsection we described an action of $T$ on the semicontinuous limit space $\mathcal{H}$. To build our bulk Hilbert space we single out the state  $[|\Omega\rangle]$ with the geometry of the standard tessellation. This is a state in the boundary Hilbert space. We build up the bulk space $\mathcal{H}_{\text{bulk}}$ by simply adjoining states to $\mathcal{H}_{\mathrm{bulk}}$ which result from the action of $f$, i.e., we set
\begin{equation}
	\mathcal{H}_{\text{bulk}} \equiv \overline{\text{span}\{\pi(f)[|\Omega\rangle]\,|\, f\in T\}}.
\end{equation}
This space is the subspace of $\mathcal{H}$ generated by all ``conformal-like'' transformations from $T$ acting on the trivial vacuum state. Because this space is closed under the action of $T$ it also gives us a unitary representation
\begin{equation}
	\pi(f)|_{\mathcal{H}_{\mathrm{bulk}}} \colon \mathcal{H}_{\mathrm{bulk}} \to \mathcal{H}_{\mathrm{bulk}}.
\end{equation}
The physical analogy to keep in mind here is that $\mathcal{H}$ should be thought of as the full CFT Hilbert space, i.e., as corresponding to the CFT vacuum plus all excitations built from the vacuum given by applying primary, secondary, tertiary fields etc. The bulk subspace  $\mathcal{H}_{\mathrm{bulk}}$ corresponds to the subspace of the full boundary CFT Hilbert space generated by the conformal vacuum plus only those states generated from the conformal vacuum by application of Virasora algebra elements, i.e., as all states built from the vacuum via local conformal transformations.

Due to the particular way Thompson's group $T$ acts we can describe an overcomplete basis of $\mathcal{H}_{\text{bulk}}$ by labelling kets with binary trees. The first observation is that the inner product between two states of the form $\pi(f)[|\Omega\rangle]$ and $\pi(g)[|\Omega\rangle]$, $f,g\in T$, in $\mathcal{H}_{\mathrm{bulk}}$ may be computed by using the group property of $T$ as follows
\begin{multline}
	(\pi(f)[|\Omega\rangle], \pi(g)[|\Omega\rangle]) = ([|\Omega\rangle], \pi(f^{-1}g)[|\Omega\rangle]) = \\ \text{some function of the reduced tree diagram $(R,S)$ for $f^{-1}g\in T$.}
\end{multline}
Using this observation we notice that we can understand all matrix elements of an operator $\pi(f)$ in $\mathcal{H}_{\mathrm{bulk}}$ in terms of vacuum matrix elements via knowledge of just $([|\Omega\rangle], \pi(h)[|\Omega\rangle])$ for all $h\in T$. The next step is to define the following special vectors in $\mathcal{H}_\gamma$: notice that any cutoff $\gamma$ determines a binary tree if we first build the partition of the unit interval and then associate to the partition to its subtree in the tree of standard dyadic intervals. If $\gamma$ has $n$ intervals this tree will have $n$ leaves; we write $R$ for this tree. Let $S$ be an arbitrary connected binary tree with $n$ leaves and define the following ket
\begin{equation}
	|R,S\rangle \in \mathcal{H}_{\text{bulk}}
\end{equation}
firstly via its inner products with kets built from other pairs of binary trees $(R,S')$ with $R$ the same but $S'$ an arbitrary binary tree with $n$ leaves according to
\begin{equation}
	\langle R,S'|R,S\rangle + \text{reflect $S$ and join its leaves to $S'$; replace vertices with $V$ and contract.}
\end{equation}
For example
\begin{equation}
	\left\langle \raisebox{-2pt}{\includegraphics{leftATree.pdf}, \includegraphics{leftATree.pdf}}\,\middle|\,\raisebox{-2pt}{\includegraphics{leftATree.pdf}, \includegraphics{rightATree.pdf}}\right\rangle \quad =\quad \raisebox{-10pt}{\includegraphics{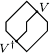}}.
\end{equation}
This definition may be extended to give the inner product between arbitrary kets $|R,S\rangle$ and $|R',S'\rangle$: the only difference is that if $R\neq R'$ we first simultaneously subdivide the pairs of trees $(R,S)$ and separately the pair $(R',S')$ by adding carets to $S$ and $R'$ until they become equal.

It is possible to argue that the kets $|R,S\rangle$ defined by pairs of trees uniquely determine elements of $\mathcal{H}_{\text{bulk}}$, i.e., the process of adding carets is consistent with the equivalence relation $\sim$ on $\widehat{\mathcal{H}}$.

\section{The \texorpdfstring{$\mathit{Pt}$}{P}/\texorpdfstring{$T$}{T} correspondence}

We have focussed so far on building dynamics for the boundary theory of a holographic state, and what that means. We have argued that the Hilbert space of the boundary (which would be Hilbert space of the CFT part of the AdS/CFT correspondence) should be given by the semicontinuous limit and that the role of the dynamics of the boundary should be taken by Thompson's group $T$. However, we have neglected all discussion of the bulk, and what the analogous objects are here. In this section we argue that there is an analogy between the group of (large) asymptotic diffeomorphisms of the bulk on the continuous side and something called the \emph{Ptolemy groupoid} on the semicontinuous side. The object that takes the place of a quantum gravity in bulk then consists of: (1) an isometry from the subspace of ``semiclassical'' bulk states into the boundary Hilbert space (the semicontinuous limit $\mathcal{H}$) $\Phi\colon\mathcal{H}_{\mathrm{bulk}}\to \mathcal{H}_{\mathrm{boundary}} \equiv \mathcal{H}$; and (2) a unitary representation of a group of large ``discretised'' diffeomorphisms, known as the \emph{Ptolemy group} $\mathit{Pt}$, on the subspace $\mathcal{H}_{\mathrm{bulk}}$. Further, these discretised bulk diffeomorphisms will correspond precisely with the group $T$ of Thompson ``conformal'' transformations of the boundary. This is the closest analogue to the situation in the standard AdS/CFT correspondence: since small bulk diffeomorphisms are gauge transformations for the bulk the only transformations which can generate physically different states are those bulk diffeomorphisms with a nontrivial asymptotic action, in which case the bulk diffeomorphisms are dual to their restrictions on the boundary \cite{brown_central_1986}. The results in this section build heavily on the work of Penner and coworkers, see especially the volume \cite{schneps_geometric_1997} for details.

Before we define the Ptolemy group, we first have to agree on what a ``discretised diffeomorphism'' ought to be. For this paper we think of a tessellation with distinguished oriented edge $(\tau,e)$ as defining something like a geometrical structure for $\mathbb{D}$ and hence, by specifying geodesics in the boundary of $\mathrm{AdS}_3$ via the vertices of the tessellation, for $\mathrm{AdS}_{3}$ (see \S\ref{sec:ads3} for the description of geodesics in the boundary of $\mathrm{AdS}$). If tessellations with distinguished oriented edge are like geometries then a ``discretised diffeomorphism'' ought to be a map $F$ which takes one tessellation with distinguished oriented edge $(\tau, e)$ and gives us another $(\tau', e') = F(\tau,e)$. Such a map $F$ is currently only \emph{partially defined} by this prescription: it only makes sense if you give it precisely the tessellation with distinguished oriented edge $(\tau, e)$ and is otherwise undefined. We will remedy this defect in three stages.

\begin{figure}
  \centering
  \includegraphics{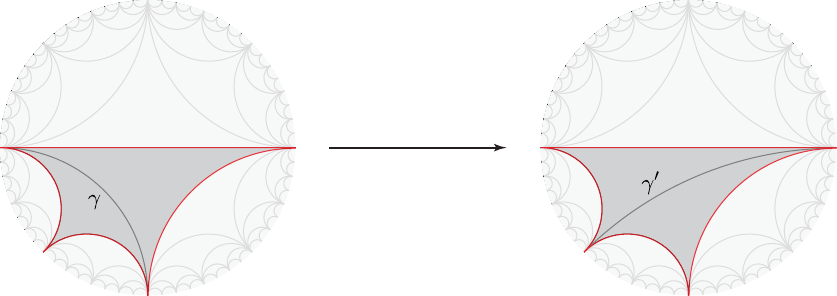}
  \caption{In a Pachner flip, or elementary move, the diagonal $\gamma$ in the shaded quadrilateral is replaced by the other diagonal $\gamma'$. The result is a new tessellation of the Poincar\'e disk.}
  \label{fig:PachnerFlip}
\end{figure}

The first stage is to agree what elementary moves we are going to allow and then build our possible maps $F$ from these moves by composition. This will already give us the structure of a groupoid (which is just a fancy word for a set with only a \emph{partially defined} group product structure). We've already identified a special move, namely, the \emph{Pachner flip} (figure~\ref{fig:PachnerFlip}): take an edge $\gamma\in \tau$ with its two neighbouring triangles which together form an ideal quadrilateral. The edge $\gamma$ is the diagonal of the quadrilateral. The Pachner flip is then the new tessellation $\tau_\gamma$ formed by removing this diagonal and replacing it with the other diagonal $\gamma'$:
\begin{equation}
	\tau_\gamma = (\tau\cup \{\gamma'\}) \setminus \{\gamma\}.
\end{equation}
The tessellation $\tau_\gamma$ is said to have resulted from an \emph{elementary move} along $\gamma\in \tau$.

As long as $\gamma$ is not the distinguished oriented edge this definition extends to tessellations with distinguished oriented edge by simply defining the oriented edge of $\tau_\gamma$ to be that inherited from $(\tau,e)$. If $\gamma$ is the distinguished oriented edge then we take the Pachner flip to act in counter-clockwise direction. Thus a Pachner flip of the quadrilateral containing the edge $e$ as its diagonal has order $4$. We now define the Ptolemy groupoid $\text{Pt}'$ to be a collection of sets $\text{Mor}((\tau,e),(\tau',e'))$ of partially defined maps $F\colon(\tau,e) \to (\tau',e')$ between pairs of tessellations with distinguished oriented edge. The set of allowed maps between a given pair of tessellations with distinguished oriented edge is given by all finite sequences of Pachner flips which produce $(\tau',e')$ from $(\tau,e)$. Note that this is nontrivial as there can be more than one such sequence. Any allowed map in $\text{Mor}((\tau,e),(\tau',e'))$ is invertible because each of the Pachner flip moves are invertible. This is a groupoid because the product operation given by composition is, at this stage, only partially defined.

\begin{figure}
  \centering
  \includegraphics{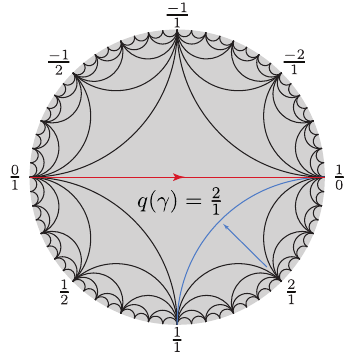}
  \caption{The labelling of the ideal vertices (endpoints of edges of the tessellation) by rational numbers.}
  \label{fig:PoincareDiscTriangulationLabelledEdge}
\end{figure}

We haven't yet solved the problem of how to independently specify a discrete diffeomorphism that can act on any input tessellation with distinguished edge $(\tau,e)$ since we have only agreed on a set of allowed maps between a given pair of tessellations with distinguished oriented edge. We want to find a recipe to select the ``same'' map from each of these sets. Thus we want to find a ``universal way'' to specify a Pachner flip which makes no reference to the tessellation it acts on. To do this realise that every edge $\gamma \in \tau$ can be specified in terms of a \emph{vertex} in the Farey tessellation: in describing Theorem~\ref{thm:penner} we built a labelling, called the \emph{characteristic map}, of the vertices of a tessellation with distinguished oriented edge via rational numbers. Our universal specification then proceeds as follows: take the unique ideal triangle containing an edge $\gamma$ which is in the component of $\mathbb{D}\setminus \gamma$ which \emph{does not} contain the distinguished oriented edge. One of the ideal vertices of this triangle does not live on $\gamma$. It is labelled by a rational number $\mathbb{Q}$ in the Farey tessellation. That rational number $q(\gamma)$ uniquely specifies $\gamma$ in $(\tau, e)$. Figure~\ref{fig:PoincareDiscTriangulationLabelledEdge} shows an example of this labelling.
Given a rational number $q$ we can now invert the process described above to uniquely specify an edge in any admissible tessellation $(\tau,e)$ with distinguished oriented edge. Suppose you have some rational number $q$ and some arbitrary admissible tessellation $(\tau,e)$, then first label all the vertices via the characteristic mapping $f_{(\tau,e)}$. Next find the vertex labelled $q$, then select the edge $\gamma$ whose label is $q(\gamma)$. As long as $q\neq\pm1$ this recipe will uniquely select an edge from $(\tau,e)$.

We now have a correspondence between $\widehat{\mathbb{Q}} \equiv \mathbb{Q} \setminus \{-1,+1\}$ and $(\tau,e)$ which identifies an edge $\gamma_q$ in $(\tau,e)$ for every $q\in \widehat{\mathbb{Q}}$. Denote by $\phi_q$, $q\in \widehat{\mathbb{Q}}$, the map that carries out the Pachner flip of the edge $\gamma_q$ in any admissible tessellation with distinguished oriented edge. There is a natural composition law for such $\phi_q$'s. The set of all such maps, and their compositions, is denoted $M$. Now two such maps $\phi, \phi' \in M$ are \emph{equivalent} if they act identically on all admissible tessellations with distinguished oriented edge. Since this is an equivalence relation $\sim$ we now only consider maps $\phi$ up to this relation $\sim$. Let $[\phi]$ denote the equivalence class of $\phi\in M$ and let $K = \{\phi\in M\,|\, \phi\sim \mathrm{id}\}$. The \emph{Ptolemy group} is now the group given by the quotient $\mathit{Pt}\equiv M/K$. It may be argued that this is a group, i.e., $K$ is big enough to allow every element to have an inverse.

It turns out that the Ptolemy group $\mathit{Pt}$ is isomorphic to none other than Thompson's group $T$. This isomorphism is not hard to guess now that we've set up all of the machinery: one has to verify that to every element $g\in \mathit{Pt}$ we can associate a unique element $f_g\in T$ (and vice versa) in a homomorphic way. The way to do this is via the characteristic mapping: an element $g\in \mathit{Pt}$ corresponds to a sequence of Pachner flips which, in turn, are associated to a homeomorphism $f_g$ of the boundary $\partial \mathbb{D} \cong S^1$.

Because the Ptolemy group $\mathit{Pt}$ acts via the same Pachner flips as $T$ we can directly take it to act on $\mathcal{H}_{\text{bulk}}$; we now have the strongest possible manifestation of the bulk/boundary correspondence. The ``discrete conformal group'' $T$ of the boundary is precisely the group of ``discrete diffeomorphisms'' of the bulk.

\section{Black holes and the \texorpdfstring{$\text{ER}=\text{EPR}$}{ER=EPR} correspondence}

One core limitation of the dyadic tessellation is that we cannot represent the geometries for gravitational solutions corresponding to particles. These are conical (see, e.g., \cite{matschull_black_1999} for a description) and require a more general hyperbolic tessellation. We consider the generalisation of our results to this situation in the next section. However, one setting we can discuss in the context of our $\mathit{Pt}/T$ correspondence is that of black hole solutions.

In $2+1$ dimensions there is a well-known family of black-hole solutions of Einstein's equations in the presence of a negative cosmological constant due to Ba\~nados, Teitelboim, and Zanelli. These geometries are multiply connected and may be produced from $\mathbb{D}$ by quotienting out by a discrete subgroup of the group of isometries of $\mathbb{D}$. We illustrate the simplest example here. This solution corresponds to the thermofield double (TFD) state according to the standard $\mathrm{AdS}$/CFT correspondence. Here we describe the analogue of the TFD for the $\mathit{Pt}/T$ correspondence.

\begin{figure}
  \centering
  \includegraphics{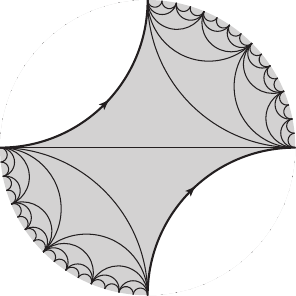}
  \caption{A tessellation $\tau_{\mathrm{BTZ}}$ for the BTZ spacetime. The two geodesics marked with arrows are identified.}
  \label{fig:BTZTriangulation}
\end{figure}

A tessellation for the BTZ spacetime can be built from the dyadic tessellation by choosing two opposite geodesics and identifying them according to the procedure outlined in \S\ref{sec:blackholes}, see figure~\ref{fig:BTZTriangulation}.
The two identified geodesics are indicated with arrows. The result of this procedure is a tessellation of the cylinder with two boundaries $A$ and $B$ at spatial infinity. The two boundaries may each be identified with $S^1$.

\begin{figure}
  \centering
  \includegraphics{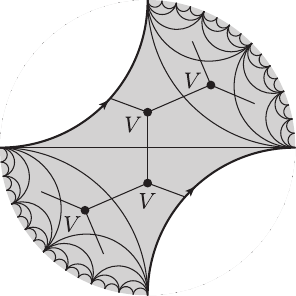}
  \caption{A tensor network for the BTZ black hole.}
  \label{fig:BTZTriangulationTNS}
\end{figure}

By associating the perfect tensor $V$ to each triangle in the BTZ tessellation we obtain the tensor network shown in figure~\ref{fig:BTZTriangulationTNS}.
This network can be thought of in two ways. Firstly, it may be understood as a state $|\Phi_{AB}\rangle$ with the geometry $\tau_{\mathrm{BTZ}}$: here we are thinking of the Hilbert space of the entire system given by $\mathcal{H}_{AB}\cong \mathcal{H}_A\otimes \mathcal{H}_B$, the semicontinuous limit built on the two boundaries $A$ and $B$ at spatial infinity. Note that $|\Phi_{AB}\rangle$ is \emph{not} a product state with respect to the tensor product over $A$ and $B$, it is an \emph{entangled state}. This gives rise to the second interpretation of $|\Phi_{AB}\rangle$, namely, as an entangled state of the two distinct subsystems $A$ and $B$. This equivalence between entanglement and geometry is the manifestation of the ER=EPR proposal \cite{maldacena_cool_2013} for the $\mathit{Pt}/T$ correspondence.

\section{Generalisations}

In this paper, we have detailed the construction of the semicontinuous limit, and dynamics thereof, for the specific case of triangular tessellations. There are now different directions for generalisation, each coming with different challenges.

One of the most important underlying features of the triangular tessellations we use is that the constructed tensor networks have a tree-like structure. This tree-like structure is the main reason it is possible to take the semicontinuous limit. Therefore, a first step would be to consider other tessellations whose underlying graph structure is that of a tree. The analogues of the modular group from our example are then given by certain Coxeter groups which are isomorphic to modular groups over extended integer rings. It is not yet known what the analogue of Thompson's group would be in this case, but a starting point might be to find a generalisation of the Ptolemy group, whose definition appears much simpler.

\begin{figure}
  \centering
  \includegraphics{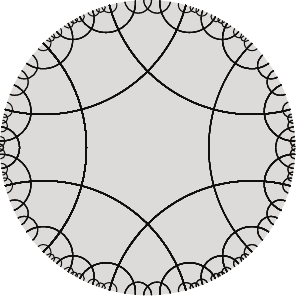}
  \caption{The hexagonal tiling $\{6, 4\}$.}
  \label{fig:HexagonalTessellation}
\end{figure}

In general, for arbitrary regular tilings of the Poincar\'e disk the symmetry group will be some Fuchsian group. For example, in the case of a hexagonal tiling with four hexagons meeting at a vertex (figure~\ref{fig:HexagonalTessellation}) it turns out that using a 6-leg perfect tensor will give us a unitary representation of the isometry group of this tessellation, namely, a Fuchsian group $G$ given by a subgroup of the quaternions $\mathbb{Q}$ defined over the field $\mathbb{Q}[\sqrt{2}+\sqrt{3}]$.
This is not exactly a ``big'' group in the sense that Thompson's group $T$ is the group of all \emph{piecewise} $\mathit{PSL}(2,\mathbb{Z})$ functions, but it is still pretty big. The physical analogy here is that the group $G$ is a group of only global symmetries, whereas Thompson's group $T$ is analogous to the conformal group which is a group of local symmetries. We very nearly do get unitary representations of an analogous group of piecewise $G$ transformations, but this just fails (it turns out that the semicontinuous limit is not continuous enough for these transformations to admit unitary representations).

In physical terms what this means is that there is an analogue of the group of global conformal transformations which acts unitarily, but that local conformal transformations do not directly act unitarily on the semicontinuous limit Hilbert space unless we either: (1) find special perfect tensors that satisfy additional constraints; or (2) we augment the semicontinuous limit Hilbert space by adding in states via a process known as completion. The second option is not entirely desirable as it is completely unclear whether the resulting Hilbert space is separable.

\begin{figure}
  \centering
  \includegraphics{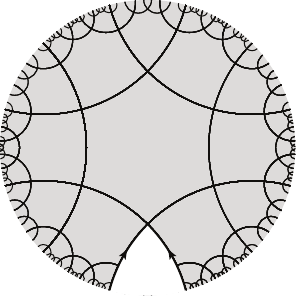}
  \caption{The hexagonal tiling $\{6, 4\}$ with a wedge cut out. The two geodesic segments marked by arrows are identified. We propose this as a model for a massive particle solution.}
  \label{fig:HexagonalTessellationParticle}
\end{figure}

One benefit to considering more general tessellations such as the $\{6, 4\}$ tessellation here is that we can represent bulk particle solutions. This is easy to do by forming conical geometries by deleting and gluing. For example, figure~\ref{fig:HexagonalTessellationParticle} shows a massive particle solution built by cutting out a quarter of the tessellation and gluing the exposed edges.
By substituting a $6$-leg perfect tensor in place of every polygon in this tessellation and contracting we obtain a semicontinuous Hilbert space etc. We also get a unitary action of a subgroup of the group $G$.

Finally, regarding holographic \emph{codes} instead of states, there is really no work to do: the semicontinous limit can be constructed in the same way as before since additional bulk legs do not interfere with the procedure. In the limit we then obtain a space of linear maps (instead of a Hilbert space) containing holographic codes; Thompson's group $T$ acts in exactly the same way as before.

\section{Conclusions and outlook}

In this paper we have commenced the construction of a dynamical toy model of the AdS/CFT correspondence built on holographic codes. Our findings are summarised in table~\ref{tab:dict}.
Many obvious steps remain incomplete, including, the discussion of fields and their holographic duals, the Ryu-Takayanagi formula, and others.

\begin{table}
	\centering
	\begin{tabular}{c|c}
		\hline
		Continuum & Discretuum \\
		\hline
		Poincar\'e disk $\mathbb{D}$ & Tessellation $(\tau,e)$ \\
		$\conf(\mathbb{R}^{1,1})$ & $T\times T$ \\
		CFT Hilbert space $\mathcal{H}_{\mathrm{CFT}}$ & semicontinuous limit $\mathcal{H}$ \\
		$\mathcal{H}_{\mathrm{CFT}} \subset \mathcal{H}_{\mathrm{AdS}_3}$ & $\mathcal{V}\subset \mathcal{H}$ \\
		(Large) bulk diffeomorphisms & Pachner flip \\
		Group of bulk diffeomorphisms & Ptolemy group $\mathit{Pt}$
  \end{tabular}
  \caption{Holographic dictionary for the $\mathit{Pt}/T$ correspondence}
  \label{tab:dict}
\end{table}

\subsection{Fields for Thompson's groups}
We have argued that there is a strong analogy between the conformal group and Thompson's group $T$. One might hope that this analogy is not accidental and can be extended to build a theory of Thompson-symmetric fields in parallel with conformal field theory. It turns out that one \emph{can} define analogues of primary field operators for Thompson's group $T$ and that they transform in an analogous way to their conformal counterparts under Thompson group transformations. Additionally, these field operators enjoy fusion rules and give rise to descendant fields as one might hope \cite{osbornestiegemann2019}.

\subsection{Bulk fields}
It is straightforward to add bulk fields by exploiting \emph{pluperfect} tensors \cite{yang_bidirectional_2016}. This should give rise to something like an intertwiner for Thompson group representations. We have not explored this in any detail.

\subsection{More general tessellations and MERA}
Very recent work of Evenbly \cite{evenbly_hyper-invariant_2017} has lead to the construction of MERA-like networks with perfect-like properties. These seem a particularly promising place to look for Thompson-like discrete local symmetry groups which are not based on trees.

\acknowledgments

Firstly, we'd like to sincerely thank Vaughan Jones and Yunxiang Ren for many helpful discussions and for extremely valuable guidance, especially with the mathematical theory of subfactors, planar algebras, and the Thompson group. We are also grateful to C\'edric B\'eny, Gemma De las Cuevas, Robert K\"onig, Fernando Pastawski, and Mario Szegedy for numerous comments and suggestions. Finally, we thank the anonymous referees of the Quantum Information Processing (QIP) conference 2018 in Delft for their helpful reports.

This work was supported by the DFG through SFB 1227 (DQ-mat) and the RTG 1991, the ERC grants QFTCMPS and SIQS, the cluster of excellence EXC201 Quantum Engineering and Space-Time Research, and the Australian Research Council Centre of Excellence for Engineered Quantum Systems (EQUS, CE170100009).

\bibliographystyle{unsrt}

\bibliography{lit}

\end{document}